\documentclass[graybox, envcountchap]{svmult}

\usepackage{mathptmx}        
\usepackage{helvet}          
\usepackage{courier}         

\usepackage{makeidx}         
\usepackage{graphicx}        
\usepackage{multicol}        
\usepackage[bottom]{footmisc}

\makeindex             

\usepackage{amsmath}
\usepackage{amssymb}
\usepackage{lscape}
\usepackage{multirow}

\def\Msun{{\rm M}_\odot}

\def\gapp{\ifmmode\stackrel{>}{_{\sim}}\else$\stackrel{<}{_{\sim}}$\fi}
\def\gsim{\lower.5ex\hbox{\gtsima}}
\def\gtsima{$\; \buildrel > \over \sim \;$}
\def\kms{\rm km~s$^{-1}$}
\def\lapp{\ifmmode\stackrel{<}{_{\sim}}\else$\stackrel{<}{_{\sim}}$\fi}
\def\lsim{\lower.5ex\hbox{\ltsima}}
\def\ltsima{$\; \buildrel < \over \sim \;$}

\newcommand\apgt{\ {\raise-.5ex\hbox{$\buildrel>\over\sim$}}\ }
\newcommand\aplt{\ {\raise-.5ex\hbox{$\buildrel<\over\sim$}}\ }
\begin{document}
\pagestyle{empty}
\frontmatter


\mainmatter

\setcounter{chapter}{2}

\title{The Blue Stragglers of the\\ Old Open Cluster NGC 188}
\author{Robert D. Mathieu \and Aaron M. Geller}
\institute{Robert D. Mathieu  \at Department of Astronomy, University of Wisconsin Ð Madison, Madison WI 53726 USA\\
 \email{mathieu@astro.wisc.edu} \\
 \and Aaron M. Geller \at Department of Physics and Astronomy, Northwestern University, 2145 Sheridan Road, Evanston, IL 60208, USA,
 \email{a-geller@nortwestern.edu}  }

%
%
\maketitle
\label{Chapter:Mathieu}

\abstract*{The old (7 Gyr) open cluster NGC 188 has yielded a wealth of astrophysical insight into its rich blue straggler population. Specifically, the NGC 188 blue stragglers are characterised by:
\begin{itemize}
\item A binary frequency of 80\% for orbital periods less than $10^4$ days;
\item Typical orbital periods around 1000 days;
\item Typical secondary star masses of 0.5 $\Msun$;
\item At least some white dwarf companion stars;
\item Modestly rapid rotation;
\item A bimodal radial spatial distribution;
\item Dynamical masses greater than standard stellar evolution masses (based on short-period binaries);
\item Underluminosity for dynamical masses (short-period binaries).
\end{itemize}
Extensive $N$-body modeling of NGC 188 with empirical initial conditions reproduces the properties of the cluster, and in particular the main-sequence solar-type binary population. The current models also reproduce well the binary orbital properties of the blue stragglers, but fall well short of producing the observed number of blue stragglers. This deficit could be resolved by reducing the frequency of common-envelope evolution during Roche lobe overflow.
Both the observations and the $N$-body models strongly indicate that the long-period blue-straggler binaries --- which dominate the NGC 188 blue straggler population --- are formed by asymptotic-giant (primarily) and red-giant mass transfer onto main sequence stars. The models suggest that the few non-velocity-variable blue stragglers formed from mergers or collisions. Several remarkable short-period double-lined binaries point to the importance of subsequent dynamical exchange encounters, and provide at least one example of a likely collisional origin for a blue straggler.
}

\label{matsec1}
\section{Blue Stragglers in Open Clusters}
\label{matsec11}

While the photometric study of the globular cluster M3 \cite{Sandage1953} is often cited as a seminal introduction of blue stragglers, the demonstration of blue stragglers in the open cluster\index{open cluster} M67\index{M67} comes soon thereafter in the classic paper of \cite{JSandage1955}.

Since then blue stragglers have been found in large numbers of open clusters. For example, Ahumada and Lapasset  \cite{AL2007} identify 1887 blue straggler candidates in 427 open clusters, and provide a catalog with a wide array of data for these blue straggler candidates. Ahumada and Lapasset \cite{AL1995} did macroscopic analyses searching for trends across the open cluster blue straggler populations (using an earlier version of the catalogue), to which the reader is recommended.

As often in science, understanding is gained by combining both macroscopic and microscopic perspectives. Since the discovery paper of \cite{JSandage1955}, M67 has been a primary laboratory for detailed and fruitful studies of a specific set of blue stragglers (e.g. \cite{Mathys1991}, \cite{MathieuLatham1986}, \cite{LeonardLinnell1992}, \cite{Latham2007}).

Recently we have developed a new laboratory, the old open cluster NGC 188\index{NGC 188}. The 20 blue stragglers of this cluster have proven a rich vein that we have mined deeply. Here we first present a comprehensive observational picture of these blue stragglers, followed by comparison of the actual blue stragglers with the products of $N$-body\index{N-body simulation} simulations of NGC 188. Ultimately we suggest that most of the blue stragglers in NGC 188 are the result of mass-transfer\index{mass transfer} processes, quite likely joined by contributions from both collision and merger formation channels.

\begin{figure}
\includegraphics[width=119mm]{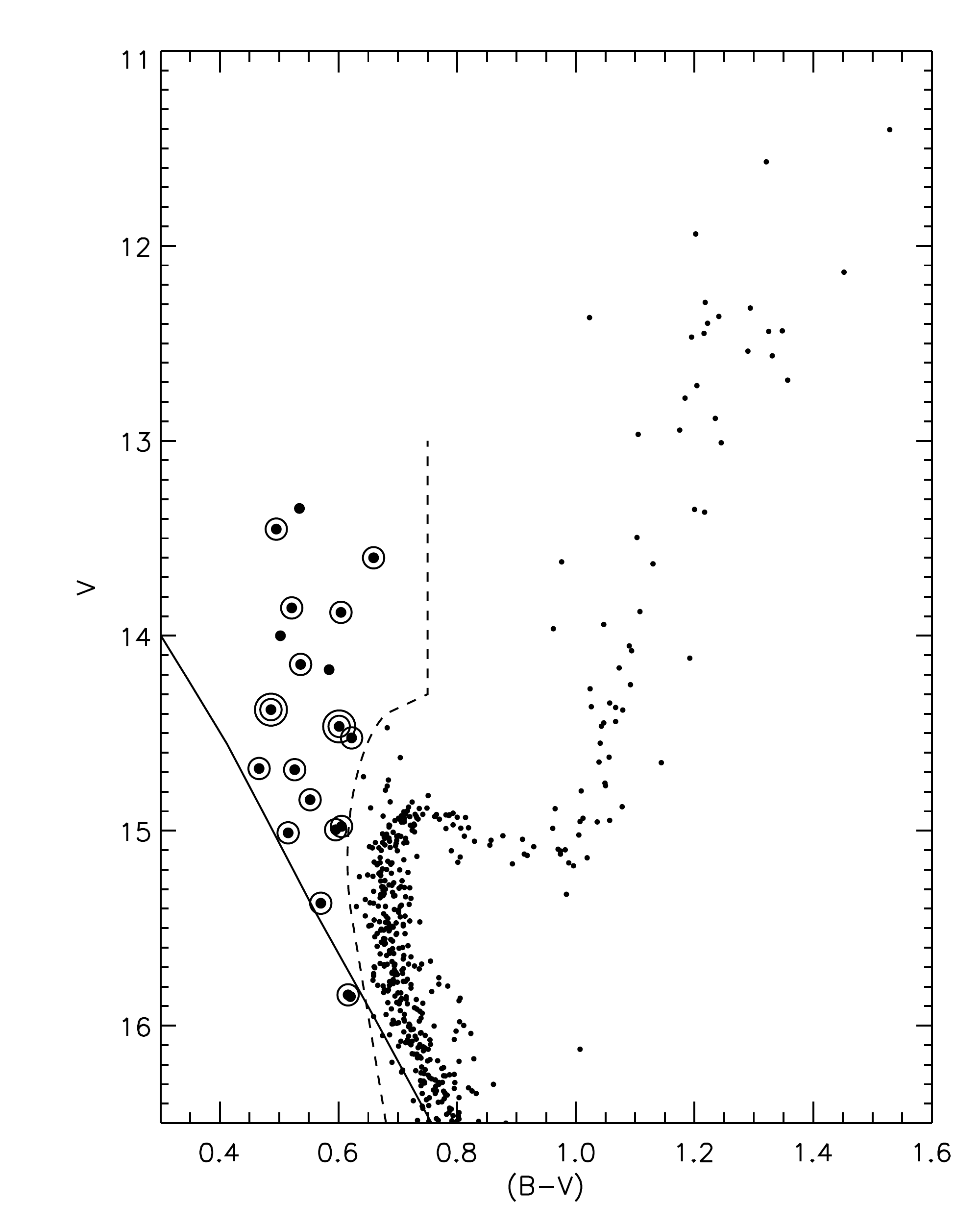} 
\caption{Colour-magnitude diagram\index{colour-magnitude diagram} for NGC 188. The 20 blue stragglers are shown as larger black dots and are identified as being to the left of the dashed line at a given magnitude. The line is drawn so as to not include binaries comprising two normal stars from the cluster turn-off region. For reference, we also show the zero-age main sequence (solid line). Updated from \cite{MathieuGeller2009}.}
\label{matfig1}
\end{figure}

\subsection{The Open Cluster NGC 188}
\label{matsec12}

The open cluster NGC~188 is one of the richest old open clusters in the Galaxy. A recent proper-motion study finds that NGC~188 contains about 1500 cluster members down to a magnitude of $V = 21$ (0.5 $\Msun$) within a 17 pc radius (13 core radii) in projection \cite{Plataisetal2003}. Its old age (7 Gyr), reasonable proximity to the Sun (2 kpc) and low reddening ($E(B-V)=0.09$; distance and reddening from \cite{Sarajedinietal1999}) have attracted observational studies for more than 50 years. Together these have yielded a wealth of astro-physical data that provides a broad foundation from which to study a dynamically evolved open cluster.

NGC 188 also includes a rich population of 20 blue stragglers\footnote{Our previous papers identified 21 blue stragglers. Recently we found an error in the photometry for WOCS ID 1947 in \cite{Plataisetal2003}. In fact star 1947 is a giant; data from the AAVSO Photometric All Sky Survey (APASS) yield $V$ = 12.54 and $(B-V)$ = 1.29.  We have updated appropriately all of the figures and accounting here.  We also note that there are two proper-motion\index{proper motion} members (WOCS IDs 4230 and 4447 from \cite{Plataisetal2003}) in the blue-straggler region of the colour-magnitude diagram for which we are unable to derive precise radial-velocity measurements due to rapid rotation.  Both stars are likely members ($\geq$90\%) based on proper motions. Our mean radial velocities for both stars place them somewhat outside of the cluster velocity distribution, but we are not sufficiently confident in the radial-velocity measurement errors to report on their membership probability. WOCS ID 4230 is an X-ray source (S27, \cite{Gondoin2005}) and a radial-velocity variable. WOCS ID 4447 does not appear to be a radial-velocity variable.}, as shown in Fig.~\ref{matfig1}.  Importantly, because of the age and solar metallicity of NGC 188, these blue stragglers are in fact not blue --- they are late-F and early-G stars with effective temperatures\index{effective temperature} ranging from 6000 K to 6500 K. The significance of this lies first in permitting the measurement of highly precise radial velocities for these blue stragglers, yielding all of the results to be reported here, and then more broadly in placing the interpretation of these blue stragglers in the context of our rich understanding of the astrophysics of late-type stars.

\subsection{The WIYN Open Cluster Study and Radial Velocities}
\label{matsec13}

The WIYN Observatory 3.5-m telescope combines sizable aperture, wide field of view and excellent image quality. Instrumented with a multi-object spectrograph providing intermediate spectral resolution (R $\sim 15,000 - 20,000$) and a large format CCD imager, it is optimally suited for studies of open clusters. The WIYN Open Cluster Study\index{WIYN Open Cluster Study} (WOCS) is a broad collaboration of investigators within and beyond the WIYN consortium who together seek to 1) produce comprehensive photometric, astrometric and spectroscopic data for a select set of new fundamental open clusters, including NGC 188, and 2) address key astrophysical problems with these data \cite{Mathieu2000}.

The primary role of the WOCS team at the University of Wisconsin--Madison has been to acquire time-series, precise ($\sigma$ = 0.4 \kms) radial-velocity\index{radial velocity} measurements for nearly complete samples of solar-type main-sequence stars, giants, and blue stragglers in each of the WOCS clusters. Specifically, radial-velocity measurements are obtained for well-defined samples of stars having $(B-V)_o > 0.4$ and $V < 16.5$ within 30 arcmin of the cluster's centre.

For NGC 188 specifically, the availability of the WOCS deep wide-field proper-motion\index{proper motion} study allowed straightforward definition of a sample of 1498 stars satisfying these criteria. To date we have 10,046 radial-velocity measurements of 1151 stars over 39 years, including at least three radial-velocity\index{radial velocity} measurements\footnote{We gratefully acknowledge the extensive radial-velocity observations of Hugh Harris and Robert McClure and their collaborators Roger Griffin and James Gunn, who from 1973 through 1996, executed a radial-velocity survey of 77 stars in NGC 188 and kindly merged their data with ours.} for essentially all of the 630 stars with non-zero proper-motion membership probabilities and $V \leq16.5$.  Most of these data can be found in \cite{Gelleretal2008}.

All 20 blue stragglers yield precise radial-velocity measurements, and each has been observed at least 6 times over at least 3.3 years. The discoveries from these data comprise the first half of this chapter.

\section{Observational Findings from the Blue Stragglers in NGC 188}
\label{matsec2}

\subsection{Binary Frequency}
\label{matsec21}

Remarkably, the binary frequency ($P < 10^4$ days)\footnote{Our Monte Carlo analysis indicates a detection limit for binaries of $10^4$ days, and a completeness limit for binary orbital solutions of 3000 days \cite{GellerMathieu2012}.}  among the 20 blue stragglers is 80\% $\pm$ 20\%. This binary frequency is more than three times larger than the NGC 188 solar-type main-sequence binary frequency of 23\% in the same period range \cite{MathieuGeller2009,GellerMathieu2012}.

Only four of the blue stragglers are not detected as spectroscopic binaries\index{spectroscopic binary}. Our incompleteness studies indicate a 24\% chance of having one undetected binary with a period less than $10^4$ days, with the incompleteness function rising rapidly toward $10^4$ days. Thus it is possible that some of these non-velocity-variable blue stragglers are long-period binaries, perhaps even in this period regime.

All of the blue-straggler spectroscopic binaries now have yielded orbital solutions\footnote{One of the orbital solutions has not been previously published. The orbital parameters for WOCS ID 8104 are given in Table~\ref{matwocs}}.  Two are double-lined with very short periods, to which we will return.

\begin{table}[t]
\caption{Orbital parameters for WOCS ID 8104}
\label{matwocs}
\begin{tabular}{p{5cm}p{5cm}}
\hline\noalign{\smallskip}
Parameter &	Value\\
\noalign{\smallskip}\svhline\noalign{\smallskip}
$P$ (days) 	& 2140 $\pm$ 110\\
$\gamma$ (\kms) &	$-41.2 \pm$ 0.3\\
$K$ (\kms) &	5.5 $\pm$ 0.4\\
$e$ 	& 0.20 $\pm$ 0.07\\
$\omega$ (deg) &	267 $\pm$ 21\\
$T0$ (HJD--2 400 000 d) 	& 55,390 $\pm$ 150\\
$a \sin{i} (10^6$ km) 	&178 $\pm$ 12\\
$f(m) (\Msun)$	& $ 0.039 \pm 0.008  $\\
$\sigma$ (\kms) 	& 0.7\\
$N$ &	19\\
\noalign{\smallskip}\hline\noalign{\smallskip}
\end{tabular}
\end{table}%

\begin{figure}[b]
 \includegraphics[width=119mm]{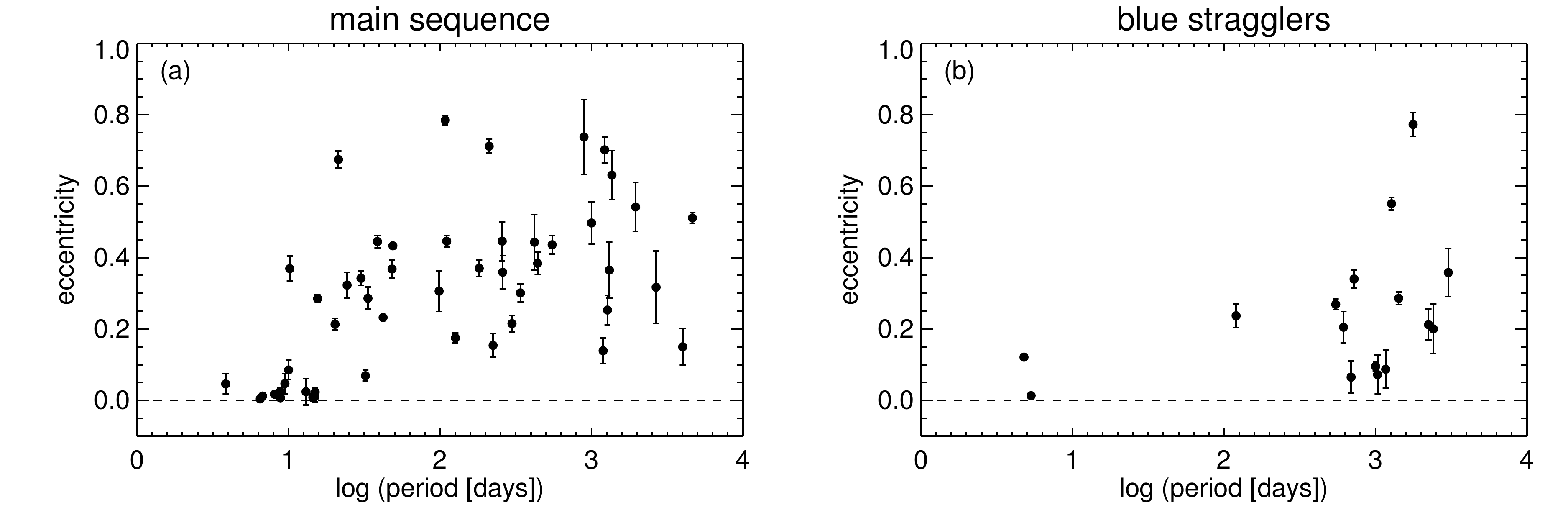}
\caption{Plots of orbital eccentricity against log period for the NGC 188 solar-type main-sequence binary stars (left panel) and for the blue-straggler binary stars (right panel). Updated from \cite{MathieuGeller2009}.}
\label{matfig2}
\end{figure}

\subsection{Orbital Period and Eccentricity Distributions}
\label{matsec22}

The distributions of two astrophysically important orbital parameters --- period\index{orbital period} and eccentricity\index{eccentricity} --- are often displayed in an $e -\log P$ diagram, as shown in Figure~\ref{matfig2} for both the solar-type main-sequence binary stars and the blue-straggler binary stars. 

The period distribution of the blue-straggler binaries is notably different from that of the main-sequence binaries. Most of the blue-straggler binaries have orbital periods within a half-decade of 1000 days, whereas the main-sequence binaries in NGC 188 (and in other open clusters and the Galactic field) populate all periods from this period domain down to only a few days.

Comparison with the longest period binaries detected among the NGC 188 solar-type stars --- which certainly extend to longer periods (e.g., \cite{Raghavan2010}) --- confirms that the upper limit on the blue straggler orbital periods is likely an observational limit. Again, some of the four ``single«« blue stragglers may have companions at longer periods.

The eccentricity distribution of the long-period ($P > 100$ days) blue stragglers also differs from the solar-type binaries; the mean eccentricity for the long-period blue-straggler binaries is 0.27 $\pm$ 0.05, while the mean eccentricity for similar main-sequence binaries is higher at 0.42 $\pm$ 0.04. The formal likelihood of the two distributions being the same is $< 1$\% (updated from  \cite{GellerMathieu2012,MathieuGeller2009}). 

Of special physical significance are the three long-period binaries with eccentricity measurements consistent with circular orbits. None of the solar-type main-sequence binaries with periods longer than the tidal circularisation\index{tidal circularisation} period of 14.5 days \cite{MeibomMathieu2005} show such low orbital eccentricities, and they are similarly rare among solar-type binaries in other clusters and the field. Tidal circularisation at such long periods would strongly suggest the earlier presence of an evolved star --- and possibly mass transfer\index{mass transfer} --- in each of these blue straggler binaries.

Even so, most of the long-period blue-straggler binaries do have substantial orbital eccentricities, which as we will see later pose an interesting challenge to mass-transfer scenarios for the formation of blue stragglers.

The two double-lined binaries are also notable for their uniquely short periods (4.8 days and 5.3 days) among the NGC 188 blue stragglers (Figure~\ref{matfig2}), which later in the chapter we attribute to dynamical encounters for at least one if not both cases. In addition, WOCS ID 5078 has a non-zero orbital eccentricity of 0.12, despite having an orbital period of 4.8 days, well below the tidal circularisation period of 14.5 days. Interestingly, the blue-straggler binary F190 in M67 has remarkably similar orbital properties \cite{MiloneLatham1992}. 

Eccentricities of these small magnitudes could be excited by an as yet undetected tertiary companion. Non-zero orbital eccentricities are also expected for the products of stellar dynamical encounters\index{stellar encounter}, although higher eccentricities are favoured. Interestingly, such a dynamical origin would require the encounter to have been more recent than the short tidal circularisation timescale at these small orbital periods.

\begin{figure}
\sidecaption
\includegraphics[width=119mm]{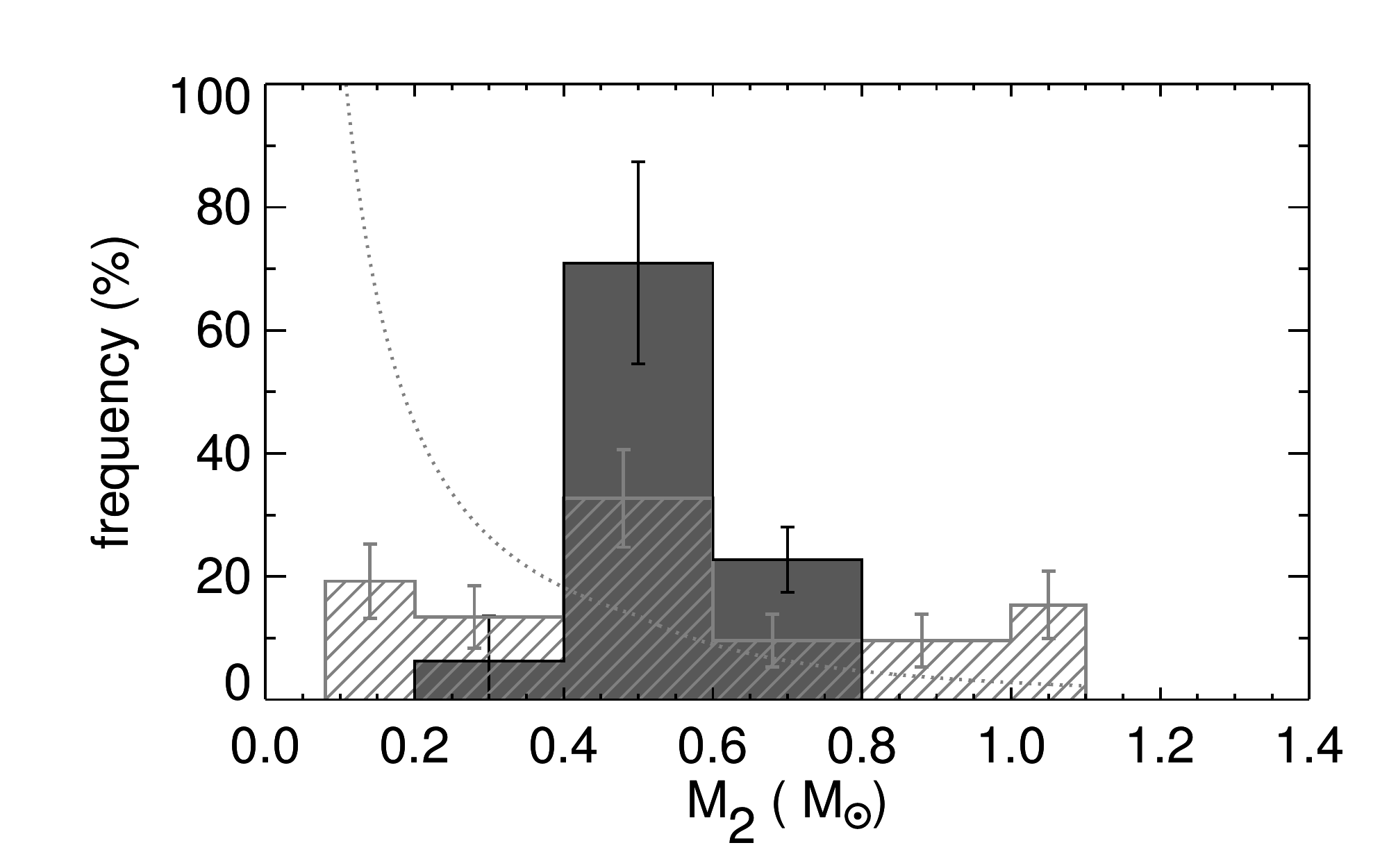} 
\caption{Secondary-star mass distribution for the 14 blue straggler binaries in NGC 188 with periods of order 1000?days. For comparison, we also plot (dashed line) an initial mass function for single stars with masses between 0.08 $\Msun$ and 1.1 $\Msun$ (from the hydrogen-burning limit to the current main-sequence turn-off mass in NGC 188). This distribution is similar to that of the secondary masses of the NGC 188 main-sequence binaries. The grey hatched histogram shows the field tertiary mass distribution, evolved to 7 Gyr in isolation (discussed in Sect.~\ref{matsec343}). Updated from \cite{GellerMathieu2011}.}
\label{matfig3}
\end{figure}

\subsection{Secondary-Star Mass Distribution}
\label{matsec23}

Only two of the blue-straggler binaries are double-lined\index{double-lined spectroscopic binary}; for the other 14 our information about the secondary stars derives only from the mass functions of single-lined orbital solutions\index{single-lined spectroscopic binary}. Given the unknown inclination, and indeed the poorly known masses of the blue straggler primary stars, standard statistical techniques \cite{MazehGoldberg1992} must be used to determine the mass distribution of the secondary stars of these blue-straggler binaries. As noted earlier, the double-lined systems are astrophysically distinct from the longer period single-lined binaries, and so we do not include them in the analysis here but discuss them separately in Sect.~\ref{matsec26}.

The secondary-star mass distribution\index{mass ratio distribution} of the long-period ($\sim$1000 days) blue-straggler binaries is shown in Figure~\ref{matfig3}. It too is noteworthy, in this case for its narrowness around a peak near 0.5 $\Msun$ \cite{GellerMathieu2011}. This distribution is markedly (and formally) distinct from the secondary mass distribution of the NGC 188 solar-type main-sequence binaries, which is consistent with a single-star initial mass function (Figure ~\ref{matfig3}; \cite{GellerMathieu2012}).

That the distribution peaks near 0.5 $\Msun$ likely has important physical significance, since this is also the expected mass of carbon-oxygen white dwarfs\index{white dwarf} left behind after mass transfer from an asymptotic giant branch star\index{asymptotic giant branch star}. 

Such asymptotic giant branch cores would not be expected to have masses lower than 0.4 $\Msun$, this being the core mass that solar-type stars develop at
the top of the red giant branch. While the distribution in Figure~\ref{matfig3} is statistical, the suggestion of one or two cases with lower secondary star masses might point toward occasional red-giant mass transfer\index{mass transfer}. 

The 120-day period of blue straggler WOCS ID 5379, not included in the
analysis for Figure~\ref{matfig3}, is particularly suggestive of a red-giant-branch mass-transfer origin. The mass function for this system is consistent with the theoretical prediction of a helium white dwarf\index{helium white dwarf} companion with a mass of 0.25 $\Msun$ to 0.5 $\Msun$ as the remnant of red-giant-branch mass transfer. \emph{Hubble Space Telescope}\index{Hubble Space Telescope} far-ultraviolet\index{ultraviolet} photometric observations have detected light from a white dwarf companion (next section), but as yet its mass is not determined. 

\begin{figure}
\includegraphics[width=119mm]{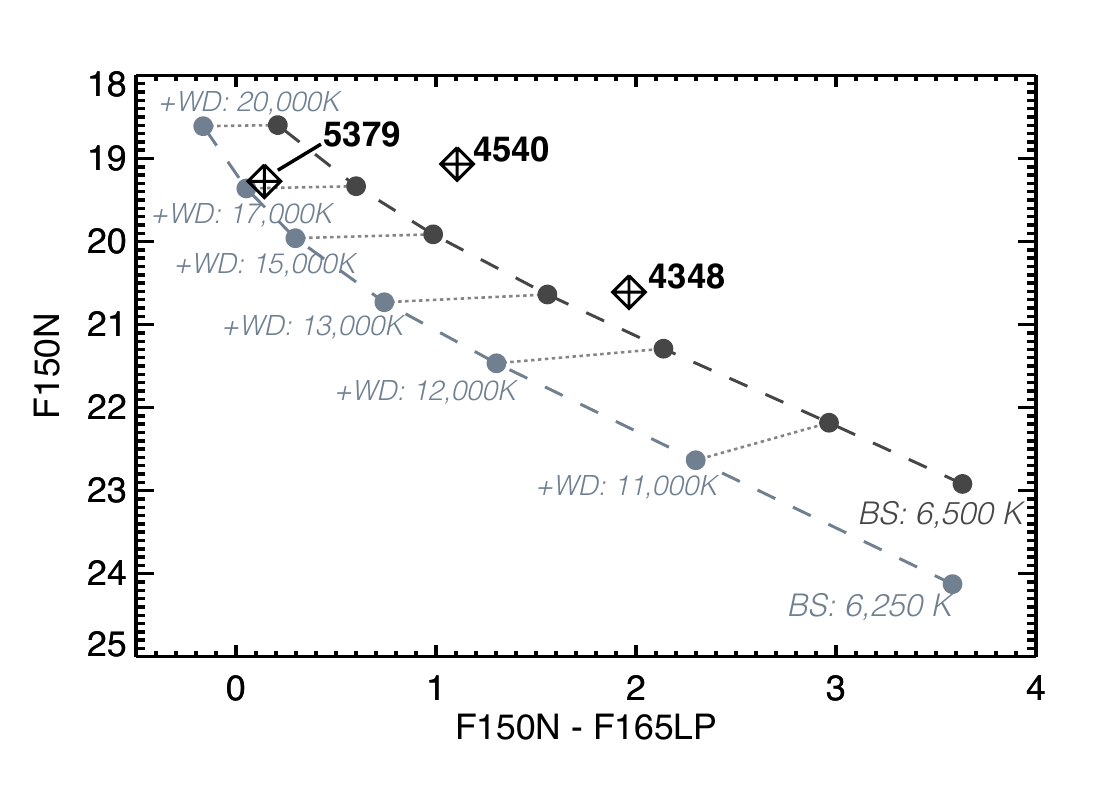} 
\caption{Far-ultraviolet colour-magnitude diagram showing the positions of three detected NGC 188 blue stragglers. F150N is a derived bandpass centered on 1554 \AA with a width of 140 \AA; F165LP is a bandpass peaking at 1700 \AA, extending from 1650 \AA with a long-pass extension to $\sim$1850 \AA. The detected blue stragglers are shown as diamonds (2$\sigma$ error bars are approximately the same size as the symbols). All three blue stragglers are also detected in F140N, a derived narrow bandpass centered on 1417 \AA. Synthetic blue-straggler -- white-dwarf binaries are shown as light grey (6250 K blue straggler) and dark grey (6500 K blue straggler) tracks with increasing temperature of white dwarf companions, as labeled \cite{Gosnelletal2013}.}
\label{matfig4}
\end{figure}

\subsection{Detection of White Dwarf Companions}
\label{matsec24}

The white dwarfs\index{white dwarf} suggested by the secondary-mass distribution are detectable at far-ultraviolet wavelengths, if they are young enough ($t < 0.4$ Gyr) and therefore hot enough ($T >$ 12,000 K) to be substantially bluer than the blue stragglers themselves.

We have observed each of the NGC 188 blue stragglers with the  \emph{Hubble Space Telescope} ACS/SBC camera. These data are currently under analysis
(N. Gosnell, PhD dissertation, in progress), but already there are three unambiguous detections of white dwarfs, as shown in Figure~\ref{matfig4}.

The observed far-ultraviolet fluxes imply white dwarf temperatures in excess of 12,000 K and thus white dwarf ages younger than $\sim$300 Myr \cite{BLR2001}. Such ages are negligibly small compared to the 7 Gyr age of NGC 188. Thus, in a mass-transfer formation scenario, that mass transfer has only very recently ended for these three blue stragglers.

\subsection{Stellar Rotational Velocities}
\label{matsec25}

\begin{figure}[!b]
\sidecaption
\includegraphics[width=75mm]{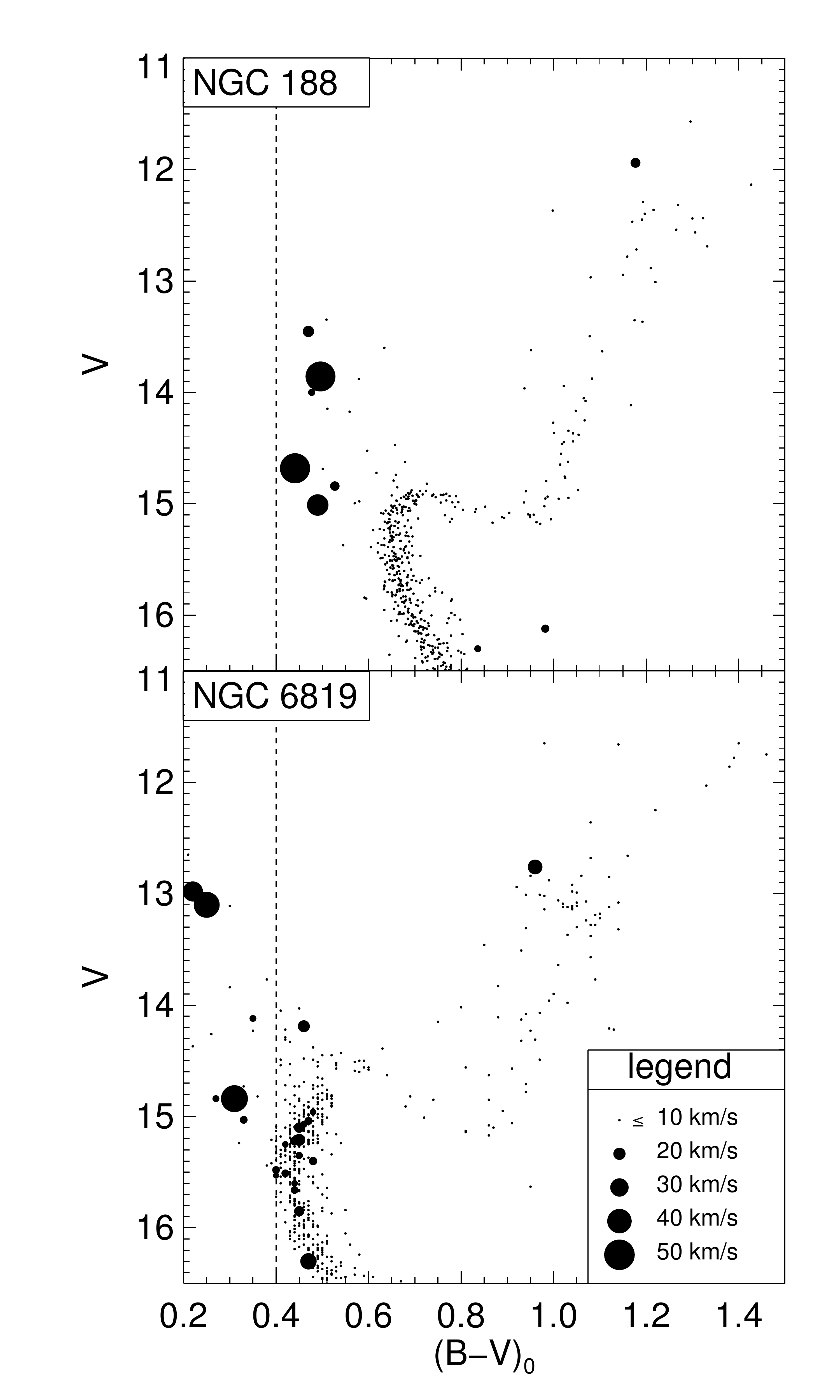} 
\caption{Colour-magnitude diagrams of NGC 188 and the 2.5 Gyr cluster NGC 6819\index{NGC 6819}, in which the point sizes correspond to the projected rotational velocities ($v \sin{i}$) for each star, as indicated. We plot only cluster members, and we have removed close binaries whose periods are less than the corresponding tidal circularisation periods, so the $v \sin{i}$ values shown have not been affected by tidal evolution. The dashed line at $(B-V)_o = 0.4$ is to guide the eye in comparing the $v \sin{i}$ values of stars of similar colours in the two clusters. Updated from \cite{MathieuGeller2009}.}
\label{matfig5}
\end{figure}

The spectra obtained for radial-velocity measurements also yield measurements of projected rotation velocities\footnote{Given our spectral resolution of 20 \kms, we only measure upper limits for stars with $v \sin{i} < 10$ \kms.}, $v \sin{i}$\index{rotation velocity}. The theoretical expectations for blue-straggler rotation velocities are not well defined, but all current formation scenarios can plausibly spin up the blue stragglers. 
In Figure~\ref{matfig5} we show the colour-magnitude diagram of NGC 188 with projected rotation velocities indicated by size of symbol. None of the main-sequence stars have measurable $v \sin{i}$, as expected for solar-type stars\index{solar-type star} at 7 Gyr. The rapidly spinning giant is a well-known FK Comae star\index{FK Comae star}. Such stars have been suggested to be recent merger\index{merger} products, perhaps from common-envelope evolution\index{common-envelope evolution}, although this scenario has challenges.

In contrast to the main-sequence stars, several of the blue stragglers show measurable rotation, in some cases higher than $v \sin{i}$ = 50 \kms. For comparison with main-sequence stars of the same effective temperature, we also show in Figure~\ref{matfig5} a similar colour-magnitude diagram\index{colour-magnitude diagram} of the 2.5 Gyr cluster NGC 6819. Although at this younger age a few of the main-sequence stars show detectable rotation, they do so at a lower frequency than among the NGC 188 blue stragglers and none have $v \sin{i}$ as large as 50 \kms.

Thus the observations show that at least some NGC 188 blue stragglers are rotating more rapidly than normal main-sequence stars at 7 Gyr, and also are more rapidly rotating than main-sequence stars of the same effective temperature at 2.5 Gyr.
From the point of view of theoretical predictions, it is not evident whether these blue stragglers are rotating fast or slow. In collision\index{collision} scenarios the more probable (high) impact parameters likely leave products with spins approaching break-up velocity\index{break-up velocity} \cite{Sillsetal2005}, and so the current rotation velocities would require a substantial spin-down mechanism. Likewise, the coalescence of two main-sequence stars is thought to lead to rapid rotation \cite{Webbink1976}. Mass transfer onto a normal main-sequence may also act to substantially spin-up the product, as predicted in certain situations \cite{Eggleton2011}.  

Interestingly, the NGC 188 blue-straggler rotation rates seem to decrease with decreasing surface temperature, as also found for normal main-sequence stars spanning the same temperature range. The spin-down of normal stars is associated with the presence of surface convective zones and magnetic fields, and the onset of stellar winds with decreasing surface temperature. The magnetised winds transport away angular momentum on timescales of several hundred million years. Whether blue stragglers of these same surface temperatures also have surface convection zones\index{surface convection zone} or magnetic fields\index{magnetic field} is unknown. If the blue stragglers do not have effective mechanisms by which to lose angular momentum, this may place serious challenges before formation mechanisms ---such as collisions and merger --- that are thought to produce very rapidly rotating products from the pre-event orbital angular momentum. Alternatively, if the internal structures, magnetic fields and winds of blue stragglers mimic normal main-sequence stars, with commensurate spin-down times, then that some blue stragglers have not yet spun down will place upper limits on their ages.

\begin{figure}
\includegraphics[width=119mm]{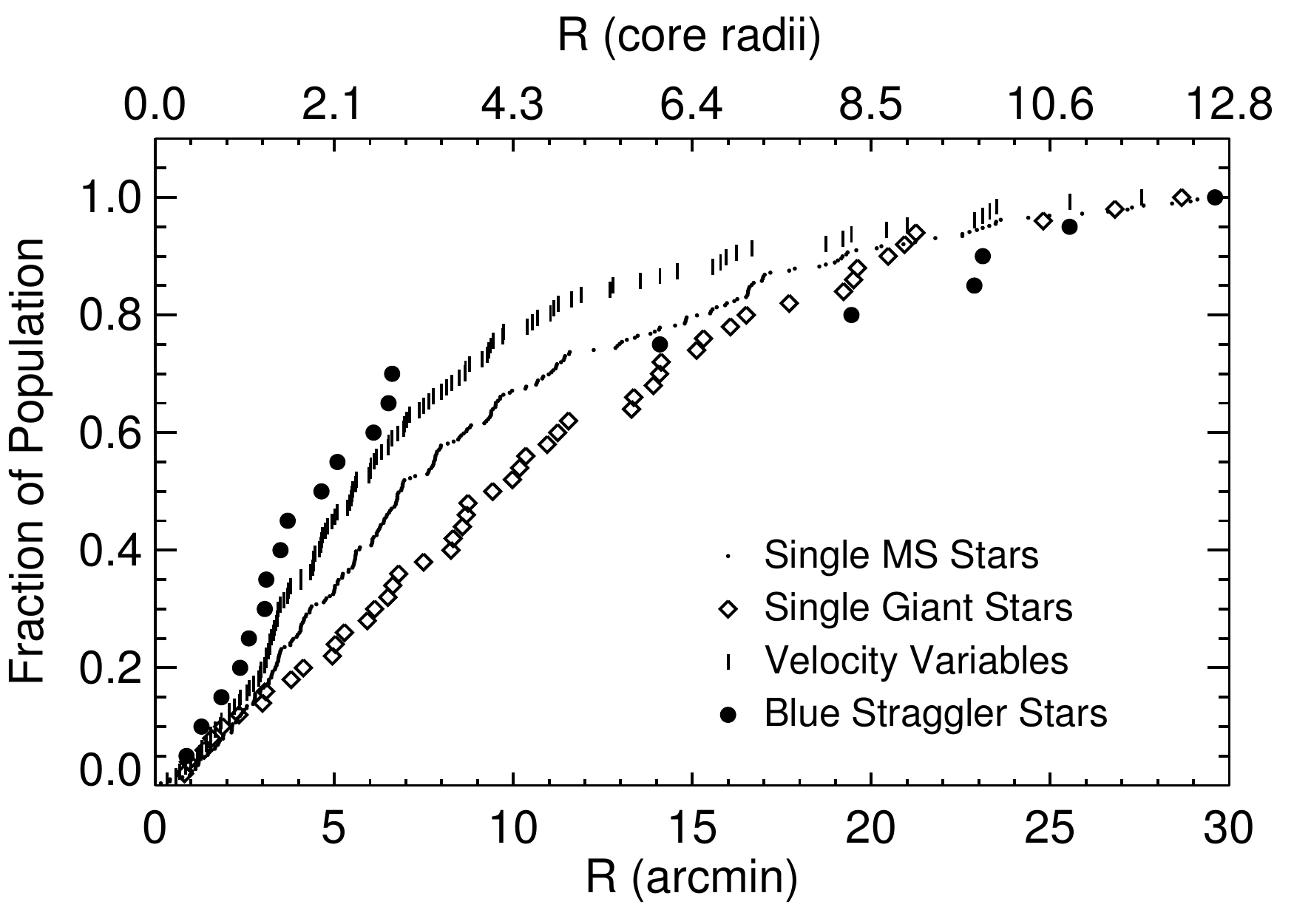} 
\caption{Cumulative spatial distribution of the single main-sequence stars, single giant stars, velocity-variable (binary) stars and blue stragglers in NGC 188. Updated from \cite{Gelleretal2008}.}
\label{matfig6}
\end{figure}

\subsection{Spatial Distribution}
\label{matsec26}

One of the more striking results of studies of globular cluster blue stragglers has been the discovery of bimodal spatial distributions\index{radial distribution} that show maxima in  blue straggler surface density, relative to normal stars, in both the cluster cores and halos, with minima in between. These have been attributed to dynamical friction causing the blue stragglers out to intermediate radii to sink to the cluster centres, while the timescales of similar processes in the halos are longer than the cluster ages (see Chap. 5).

A similar distribution of blue stragglers is found in NGC 188, as shown in Fig.~\ref{matfig6}. Note the break in the blue straggler distribution at about seven arcmin radius (three core radii). This distribution shows a centrally concentrated population, 
within about five core radii, and a halo population extending to the edge of the cluster. We discuss this distribution further in Sect. ~\ref{matsec3}.

\begin{figure}
\includegraphics[width=119mm]{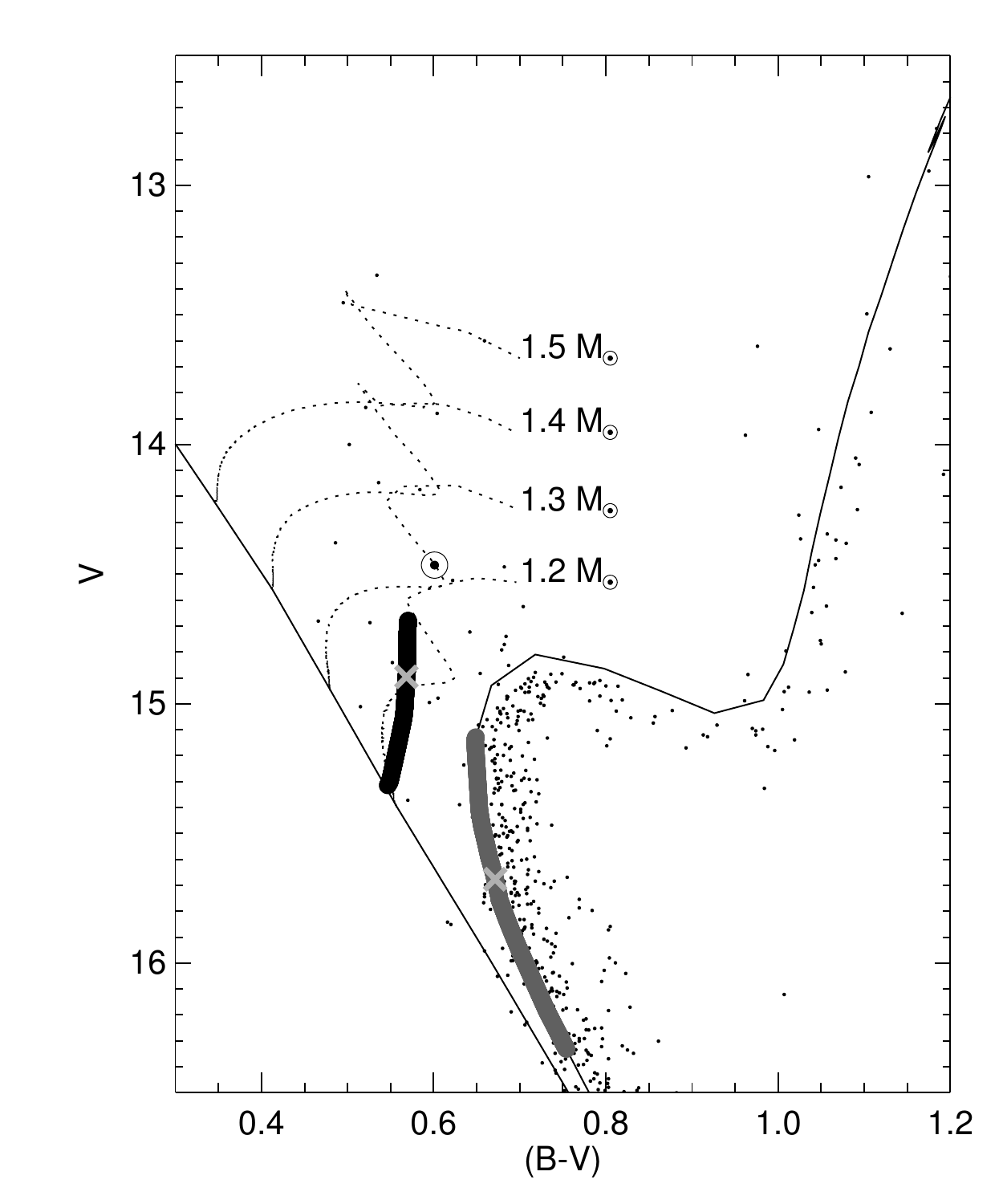} 
\caption{Colour-magnitude diagram showing the loci of possible locations for the components of the blue straggler binary WOCS ID 5078. Cluster members are shown in the small black points, and 5078 is circled. We also show a zero-age main sequence\index{ZAMS} and a 7 Gyr isochrone\index{isochrone} in the thin solid lines, and evolutionary tracks for 1.2 $\Msun$, 1.3 $\Msun$, 1.4 $\Msun$, and 1.5 $\Msun$ stars in the dotted lines, respectively \cite{Marigoetal2008}. The thick gray line shows the locus of potential secondary stars, and the thick black line shows the locus of potential primary stars. The derived effective temperature of 5850 K for the secondary implies a mass of 1.02 $\Msun$; we plot the location of a nominal secondary star, and the location of the associated primary, with the light gray crosses. Given the kinematic mass ratio of $q = 0.678 \pm 0.009$ and this secondary mass, the blue straggler mass would be 1.5 $\Msun$. Yet the blue straggler is significantly less luminous than a normal 1.5 $\Msun$ star at any point in its evolution. From \cite{GellerMathieu2012}.}
\label{matfig7}
\end{figure}

\begin{figure}
\includegraphics[width=119mm]{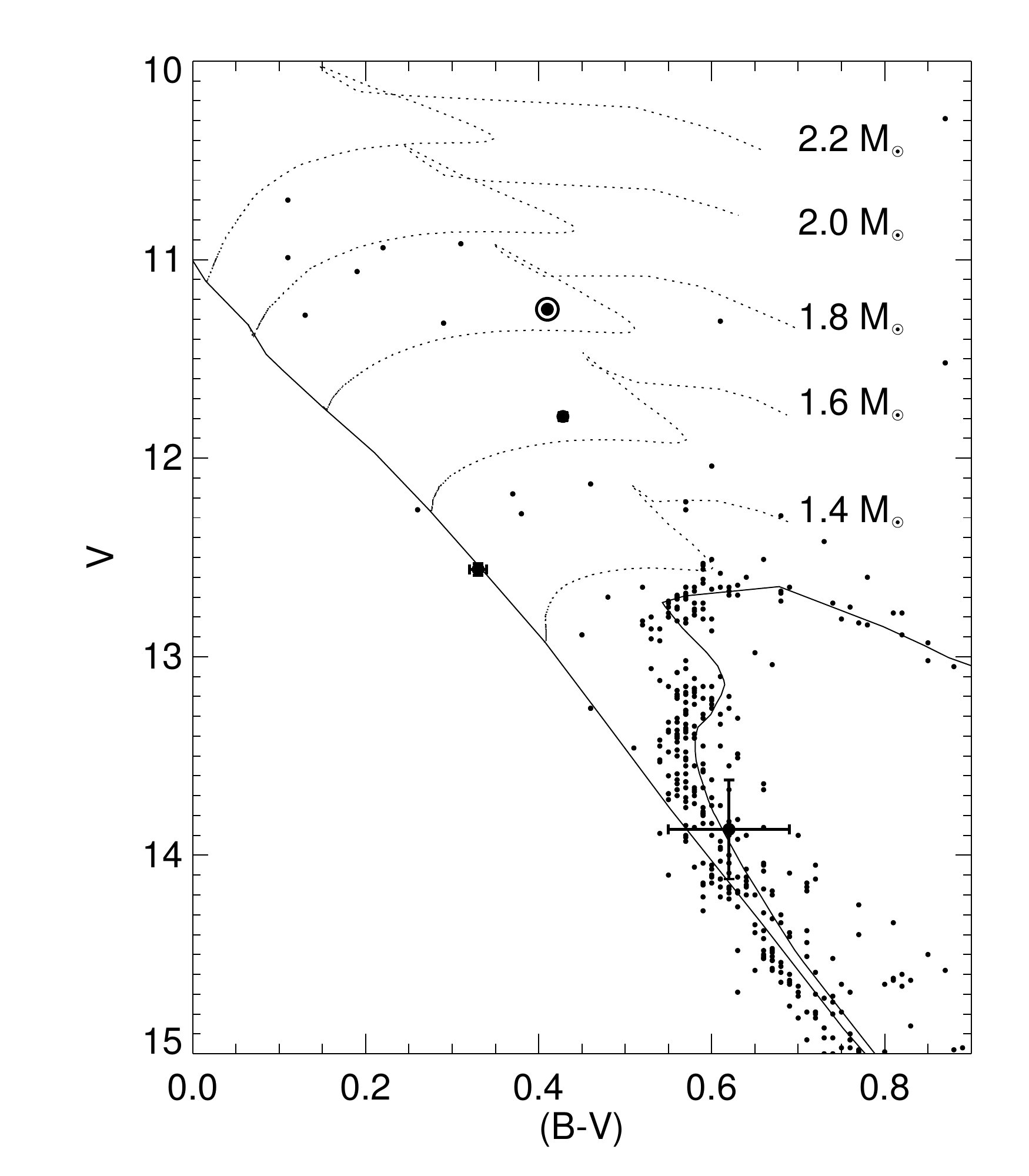} 
\caption{Colour-magnitude diagram of M67\index{M67} showing the locations of i) the combined light of blue straggler system S1082 (circled dot), ii) the tertiary blue straggler (square); and iii) the primary and secondary of the 1.07-day eclipsing binary based on the Sandquist et al. \cite{Sandquistetal2003} solution (dots with error bars). Adapted from \cite{Sandquistetal2003}.}
\label{matfig8}
\end{figure}

\subsection{Blue-Straggler Masses}
\label{matsec27}

Typically, blue-straggler masses have been derived from stellar evolution tracks for normal stars. However, the distributions of the blue stragglers in colour-magnitude diagrams do not mimic normal main-sequence stars, and in fact many do not lie between the zero-age and the terminal-age main sequences. So the appropriateness of normal stellar evolution models is not clear, and more recently researchers have sought to build evolutionary tracks that begin with initial conditions appropriate to the several current blue straggler formation scenarios, as discussed elsewhere in this book (see Chap. 12 and 14). 

Of course, what is needed are dynamical mass measurements for blue stragglers. The double-lined\index{double-lined spectroscopic binary} blue straggler binary WOCS ID 5078 takes a valuable step in this direction; its analysis is shown graphically in Fig.~\ref{matfig7}. Spectroscopic analysis of the light from the secondary star shows it to have an effective temperature of 5850 K $\pm$ 250 K (2$\sigma$). Presuming it to be a main-sequence star, the consequent mass of the secondary star is between 0.94 $\Msun$ and 1.08 $\Msun$. The mass ratio from the orbital solution is $q = 0.68$, and so the primary star mass is between 1.39 $\Msun$ and 1.59 $\Msun$. 

Figure~\ref{matfig7} shows the combined light of the system, and the deconvolved light of the blue-straggler primary star as a function of candidate main-sequence secondary stars within the permitted temperature range. Compared to a standard set of normal-star evolutionary tracks\index{evolutionary track}, the maximum inferred mass for the primary star is between 1.2 $\Msun$  and 1.3 $\Msun$. The indication then is that the normal stellar evolution models underestimate the mass of this blue straggler by roughly 15\%.

We next briefly leave NGC 188 to draw attention to a particularly important double-lined eclipsing\index{eclipsing binary} blue straggler in M67\index{M67}, S1082 \cite{vandenBergetal2001,Sandquistetal2003}. Goranskij et al. \cite{Goranskij1992} showed S1082 to have partial eclipses with a period of 1.07 days. Oddly for such a short-period eclipsing binary, radial-velocity studies of S1082 had found only very small velocity variability and no sign of a period consistent with the photometric period. The puzzle was resolved when it was shown that the system consists of (at least) four stars\index{quadruple system}, comprising both a short-period eclipsing binary and a more luminous blue straggler in a long-period binary \cite{vandenBergetal2001,Sandquistetal2003}.

Figure~\ref{matfig8} shows the colour-magnitude location of the two components of the
eclipsing binary as found\footnote{The colour-magnitude locations found for the stars in the eclipsing binary are substantially different between \cite{vandenBergetal2001} and \cite{Sandquistetal2003}, largely the result of Sandquist et al. \cite{Sandquistetal2003} including spectroscopically derived effective temperatures in their analyses. This system merits additional careful study.} by \cite{Sandquistetal2003}.  Each star currently falls near the zero-age main sequence, one as a blue straggler and one seemingly as a normal main-sequence star. Both stars become remarkable when associated with their dynamical masses. For the primary star, van den Berg et al. \cite{vandenBergetal2001} find a mass of 2.70 $\pm$ 0.38 $\Msun$  and Sandquist et al. \cite{Sandquistetal2003} find a mass of 2.52 $\pm$ 0.38 $\Msun$. As shown in Fig.~\ref{matfig8}, both of these dynamical masses greatly exceed the photometric mass of 1.5 $\Msun$ from stellar evolutionary tracks.

Even more surprising is the mass of the secondary star in the eclipsing binary\index{eclipsing binary}, which photometrically would appear to be a normal main-sequence star below the cluster turnoff. Van den Berg et al. \cite{vandenBergetal2001} find a mass of 1.70 $\pm$ 0.27 $\Msun$ and Sandquist et al. \cite{Sandquistetal2003} find a mass of 1.58 $\pm$ 0.27 $\Msun$.  However, the turnoff mass of M67 is only 1.3 $\Msun$! The expected mass for the secondary star based on the zero-age main sequence is 1.1 $\Msun$.

Formally the high mass measurement of the secondary is less than a 2$\sigma$ deviation from the zero-age-main-sequence prediction. In addition, the crucial radial-velocity measurements for this system are difficult because of both the luminous tertiary and rotational line broadening. So we must be careful that our interpretations and conclusions do not outstrip the security of the result. But the indication of very underluminous stars\index{underluminous star} in this blue-straggler binary is interesting enough to merit substantially more observational work being applied to S1082, seeking especially higher-precision mass determinations.

In closing, these systems are hopefully only beginning to tell the story of blue straggler masses. Both WOCS ID 5078 and S1082 have rather short orbital periods. There is no evidence that they are not detached binaries\index{detached binary} currently, and indeed all of the stars fit within their Roche lobes\index{Roche lobe}. Still, the implications of these findings regarding the accuracy of stellar evolution masses for blue stragglers found in wide binaries or as single stars are not yet clear. Even so, these two blue-straggler binaries present us with three underluminous stars that would seem to suggest that the standard stellar interiors theory of our undergraduate and graduate courses may not apply.

\subsection{Insights from Two Notable Blue Straggler Systems}
\label{matsec28}

The double-lined blue straggler WOCS ID 7782 may shed valuable light on the dynamical evolution of blue-straggler binaries. The key observational finding is a mass ratio of 1.005. In addition, spectroscopic analysis yields effective temperatures for the primary and secondary stars of 6500 K and 6325 K, respectively. These effective temperatures are both hotter than the main-sequence turnoff surface temperature of 5900 K. In short, this binary comprises two blue stragglers. 

No single blue-straggler formation mechanism currently under consideration is capable of forming two blue stragglers simultaneously in a 5.3-day period binary. Thus we interpret 7782 as evidence for exchanges of stars in one or more resonant dynamical encounters\index{dynamical encounter}. For specificity, consider a dynamical encounter between two of the current long-period blue straggler binaries. That such an encounter would yield a closer binary comprising the two most massive stars in the encounter (i.e., the two blue stragglers) is a probable out-come, although such a short period may be challenging (Sect.~\ref{matsec332}). 

At least one of the original companion stars would need to be ejected in order to remove energy from the system. The other companion star might do the same, or still be orbiting the blue straggler binary as an as yet undetected distant tertiary\index{triple system}.

It is also possible that only one of the initial systems contained a blue straggler, and the second one was formed by a collision during the dynamical encounter. And perhaps most likely, this system may have formed as a consequence of multiple dynamical encounters\footnote{To get a sense of the diverse dynamical stories possible for blue stragglers in clusters, the reader is recommended to the short stories told in Table 5 of Hurley et al. \cite{Hurley2002}. Case 1613 is a particular favorite of the authors.}.   

These dynamical scenarios are consistent with the fact that 7782 is in the cluster halo, 10 core radii from the centre. Momentum conservation in high-energy dynamical encounters often leads to ejections of the products from the cluster core, if not from the cluster.

Additionally, in these dynamical exchange scenarios the two blue stragglers need not have formed at the same time, and indeed their different surface temperatures might be indicative of different evolutionary ages. For example, we were able to model the system with two 1.25 $\Msun$ stars having an age difference of about two gigayears. 
Given such dynamical exchanges, the current binary system may say very little about the formation mechanism of either blue straggler. At the same time, it is worth noting that these exchange scenarios do not evidently explain the near identity of the masses of the two blue stragglers.

S1082 in M67 also leads us to consider a sequence of dynamical encounters. We introduced this four-star system\index{quadruple system} in Sect.~\ref{matsec27} in the context of the mass determinations of the eclipsing binary components; here we consider the formation of the entire system. First, we remind that the total mass of the 1.07-day period binary is measured to be 4.1--4.4 $\Msun$ (with an uncertainty of 0.5 $\Msun$). The turnoff mass of M67 is 1.3 $\Msun$. So mass contributions from at least three and likely four cluster stars are needed to create this binary. The mass of the blue-straggler primary star alone is found to be 2.5--2.7 $\Msun$, comparable to essentially the entire masses of two turnoff stars. The mass of the secondary star is found to be 1.6--1.7 $\Msun$. Clearly this blue straggler binary is not the simple product of a mass transfer event. 

In addition, the system includes another blue straggler, itself in a binary with a 1200-day period orbit. As the most luminous star in the system, it possibly is also the most massive star. The secondary star of this binary is as yet undetected in the optical. Landsman et al. \cite{Landsman1998} found S1082 to be overluminous\index{overluminous star} at far-ultraviolet\index{far ultraviolet} wavelengths and suggest the presence of a hot underluminous\index{underluminous star} companion. Should this be the dynamical companion, this long-period blue straggler binary would become intriguingly similar to the long-period blue straggler binaries in NGC 188.

Neither van den Berg et al. \cite{vandenBergetal2001} nor Sandquist et al. \cite{Sandquistetal2003} found a convincing formation path for the system, and we have done no better. We agree that the measured masses, if confirmed, strongly put forward one or more of the stars in S1082 as collision products, which in itself makes the system important as a case study. Furthermore, we see no way to form the tight binary except through a progression of dynamical encounters among multiple systems within some of which these collisions happen. Dynamical encounters favour leaving the most massive stars in the binary products, which would advantage discovery of binaries like those in S1082. Furthermore, collision products tend to be underluminous \cite{Sillsetal2001}. On the other hand, time is precious. If both stars in the close binary are highly underluminous as a result of being out of thermal equilibrium, they would remain so only for a thermal timescale. This leaves very little time, and thus likelihood, for creating the stars and the binary in more than one encounter.

If the long-period blue straggler binary in S1082 in fact is not physically associated with the close binary, then it may be a mass-transfer product much like we suggest later for the long-period binaries in NGC 188 (although it is not clear that an asymptotic giant mass transfer history could produce such a luminous blue straggler given the 2.5 $\Msun$ mass measurement of its less luminous neighbour). If they are associated, an all the more challenging series of events is required involving at least six stars. Again, the primary challenge is time, which is constrained possibly by the thermal timescales for both components of the close binary and certainly by the lifetimes of both blue stragglers.

Of course, it is puzzles such as these that make science fun, and hopefully yield unexpected discoveries. For the moment, we draw several overarching conclusions from the blue stragglers WOCS ID 7782 and S1082. First, these systems are arguably among the best observational evidence that the stellar exchanges\index{stellar exchange} in dynamical encounters\index{dynamical encounter} long predicted by theory in fact occur. Second, these systems --- as do many $N$-body simulations\index{N-body simulation} --- caution us that the properties of any given binary within which a blue straggler is found may bear little information about the formation mechanism of the blue straggler. And third, the systems --- especially S1082 --- rather strongly suggest that stellar collisions\index{stellar collision} do occur during resonant dynamical encounters of multiple systems in star clusters.

\subsection{Summary}
\label{matsec29}

In the past few years the 20 blue stragglers of NGC 188 have yielded a remarkable amount of information about the nature of blue stragglers in an old open cluster. Specifically, this blue straggler population is characterised by:
\begin{itemize}
\item A binary frequency of 80\% for P $< 10^4$ days;
\item Typical orbital periods around 1000 days;
\item Typical secondary star masses of 0.5 $\Msun$;
\item At least some white dwarf companion stars;
\item Modestly rapid rotation;
\item A bimodal radial spatial distribution;
\item Dynamical masses greater than standard stellar evolution masses (short-period binaries)
\item Underluminosity for dynamical masses (short-period binaries).
\end{itemize}

While the comprehensive scope of these results is powerful in themselves, they are further strengthened by corroborating studies. Many of these properties have also been found for the blue stragglers in the 4.5 Gyr open cluster M67 (e.g., \cite{Latham2007}). They also were foreshadowed by studies of the blue stragglers in the field (e.g.,  \cite{Carneyetal2001} and Chap. 4).

And so our microscopic perspective of the NGC 188 blue stragglers is rich with information. We now turn to the implications of this knowledge on the formation of blue stragglers. 

\section{Blue Straggler Formation Within an $N$-body Model of NGC 188}
\label{matsec3}

Current theories argue that blue stragglers form from main-sequence stars that, through interactions with an additional star or stars, gain enough mass to exceed the cluster turnoff mass.  The three currently favoured hypotheses for obtaining this extra mass are mass transfer\index{mass transfer} \cite{McCrea1964,ChenHan2008b}, mergers\index{merger} \cite{Webbink1976,ChenHan2008a} and collisions\index{collision} \cite{HillsDay1976,Sillsetal2005}.  In this section we briefly describe these formation mechanisms, and then compare with observations their predictions for the binary properties of blue stragglers based on a sophisticated $N$-body\index{N-body simulation} model of NGC 188 \cite{GellerHurleyMathieu2013}.

\subsection{The NGC 188 $N$-body Model}
\label{matsec31}

Today, $N$-body codes (e.g., {\tt NBODY6}\index{NBODY6 code}, \cite{Aarseth2003}; {\tt Starlab}\index{STARLAB code}, \cite{Hut2003}) are capable of simulating the dynamical evolution of actual open clusters from near birth across a Hubble time, including primordial populations of tens of thousands of stars and large frequencies of binaries (and triples).  Stellar evolution and detailed prescriptions for binary-star evolution are modeled self-consistently with the stellar dynamics, allowing for blue-straggler formation through all of the proposed mechanisms. 

A detailed description of the NGC 188 model can be found in \cite{GellerHurleyMathieu2013}; we briefly summarise the method here.  We use the {\tt NBODY6} code, which includes stellar and binary evolution \cite{Hurley2000,Hurley2002}, to generate 20 unique realisations of NGC 188.  For each realisation we randomly choose the initial stellar and binary parameters (e.g., positions, velocities, binary orbital periods, eccentricities, etc.) from empirically defined distributions, producing 20 unique initial stellar populations.  Each of these primordial populations is then evolved for seven gigayears. For the analyses discussed below we combine the 20 simulations to reduce stochastic effects, especially for relatively small stellar populations like blue stragglers. 

Importantly, we employ the observed solar-type main-sequence binaries of the young (150 Myr) open cluster M35\index{M35} \cite{Geller2010}  to guide our choices for the initial binary frequency and distributions of orbital elements and masses, supplemented with observations of field solar-type binaries \cite{Raghavan2010} where necessary.  These empirical initial binary conditions replace often-used theoretical distributions, which in many aspects donÕt match observed binary populations in real open clusters, both young (e.g., M35 at 150 Myr; \cite{GellerHurleyMathieu2013}) and old (e.g., NGC 188 at 7 Gyr; \cite{GellerMathieu2012}).  The importance of defining an accurate initial binary population is paramount for accurate blue-straggler production (see Sect.~\ref{matsec341}), as all viable blue-straggler formation mechanisms in open clusters begin from binary stars. 

Crucially, we find that the empirically defined initial binaries evolve within the model to reproduce well the observed NGC 188 main-sequence solar-type binary frequency and distributions of orbital parameters at seven gigayears. In addition, the NGC 188 model matches the observed cluster mass, central density and radial-density distribution.  Thus, the NGC 188 model accurately reproduces the environment in which the true NGC 188 blue stragglers formed.

Blue stragglers are identified in the model as stars that are at least 2\% more massive than the turnoff mass at a given age. We use an integrated sample of blue stragglers in our analyses of the NGC 188 model. This sample contains all blue stragglers present at each $\sim$30 Myr snapshot interval between 6 and 7.5 Gyr within all 20 simulations.  This integrated sample of blue stragglers reduces the stochastic fluctuations in blue straggler production and also weights the sample by the time that a given blue straggler spends in a specific binary, at a specific cluster radius, etc.  

\subsection{Formation Channels for Blue Stragglers in the NGC 188 Model}
\label{matsec32}

Here we briefly describe the mass-transfer, merger and collision blue-straggler formation mechanisms and their implementation in the NGC 188 model. We refer the reader to the Chap. 7, 8, 9, and 11 in this volume for more extensive discussions of these formation mechanisms, and to \cite{Hurley2002} for further details on the implementation of these formation channels in {\tt NBODY6}.

\subsubsection{Mass Transfer}
\label{matsec321}

Mass transfer through Roche Lobe Overflow (RLOF)\index{Roche lobe overflow} is typically divided into three regimes, known as Cases A, B and C, based on the evolutionary state of the donor star (\cite{KippenhahnWeigert1967}; see also \cite{Paczynski1971}). Specifically, Case A mass transfer\index{Cases A, B, C of mass transfer} occurs when the donor is a main-sequence star; Case B mass transfer occurs when the donor is on the red giant branch; and Case C mass transfer occurs when the donor is on the asymptotic giant branch. In {\tt NBODY6}, a blue straggler can be formed through mass transfer when the accretor is a main-sequence star and accepts enough material to increase its mass above that of the cluster turnoff.  If both the donor and accretor are on the main sequence, Case A mass transfer is believed to result in the coalescence (i.e. merger) of the two stars (e.g., \cite{ChenHan2008a}), and in {\tt NBODY6} binaries that come into contact are merged. Therefore, we include Case A mass transfer systems in the merger\index{merger} category, and from hereon we will use the term ``mass transfer'' to refer specifically to Cases B and C. 

In {\tt NBODY6}, RLOF transferring mass from a giant on a dynamical time-scale leads to a common-envelope\index{common-envelope evolution} episode.  It is assumed that no mass is accreted, and therefore no blue straggler is formed. As in many binary-population synthesis models (e.g., \cite{Hurley2002,Belczynsky2008}), {\tt NBODY6} employs a ``critical mass ratio'' value, $q_c$, to determine whether a given binary will undergo thermal or nuclear mass transfer (i.e., stable) or dynamical mass transfer (i.e., unstable). If $q_1 = M_{\rm donor}/M_{\rm accretor} > q_c$, the binary undergoes dynamical mass transfer, and otherwise mass transfer is stable.  

The key point here, which we will return to below, is that in {\tt NBODY6} Case B and Case C mass transfer can only produce a blue straggler if the mass transfer is stable. Given stable mass transfer, Case B mass transfer leaves a He white dwarf\index{white dwarf} companion bound to the blue straggler, while Case C mass transfer leaves a CO white dwarf companion.

\subsubsection{Mergers}
\label{matsec322}

We use the term merger\index{merger} to describe the coalescence of two members of a tight binary to form a single star. If the combined mass is large enough, a blue straggler can be produced by the merger of two main-sequence stars, which is often the result of Case A mass transfer.  Nelson \& Eggleton \cite{NelsonEggleton2001} study in detail the large variety of Case A mass-transfer scenarios.  In general, a close binary composed of two main-sequence stars can come into contact and sub-sequently merge through stellar evolution processes or due to the loss of orbital angular momentum (e.g. through magnetic braking).  

As discussed by Perets in this volume (Chap. 11), triple systems may increase the merger rate through a process involving Kozai cycles\index{Kozai cycle} and tidal friction\index{tidal friction}, known as the KCTF mechanism \cite{PeretsFabrycky2009}. The KCTF mechanism is parametrised in {\tt NBODY6} based on the analytical development of Mardling \& Aarseth \cite{Mardling2001}.  In general, if the KCTF mechanism forms a blue straggler, it will be in a binary with the original tertiary companion. 

\subsubsection{Collisions}
\label{matsec323}

We use the term collision\index{collision} to refer to a direct physical impact of two (or more) stars that subsequently combine to become a single star.   Unlike the merger and mass-transfer mechanisms, collisions require the dynamical environment of a star cluster (or a dynamically unstable multiple-star system; \cite{PeretsKratter2012}).  In open clusters like NGC 188, essentially all collisions occur during dynamical encounters involving at least one binary (or higher-order system), as the cross section for a direct single-single collision to occur is prohibitively small.

In {\tt NBODY6}\index{NBODY6 code}, a collision (or a merger) product is assumed to be fully mixed and to achieve thermal equilibrium rapidly, becoming a new main-sequence star in equilibrium upon creation \cite{Hurley2002}.  This is most likely an oversimplification, as detailed models show that collision products may not be in thermal equilibrium and furthermore may not be fully mixed \cite{Sillsetal2005,Glebbeeketal2008}.  These simplifications will affect the lifetimes and luminosities of blue stragglers formed by collisions (and mergers), but will likely not impact their binary properties.

\begin{figure}
\includegraphics[width=119mm]{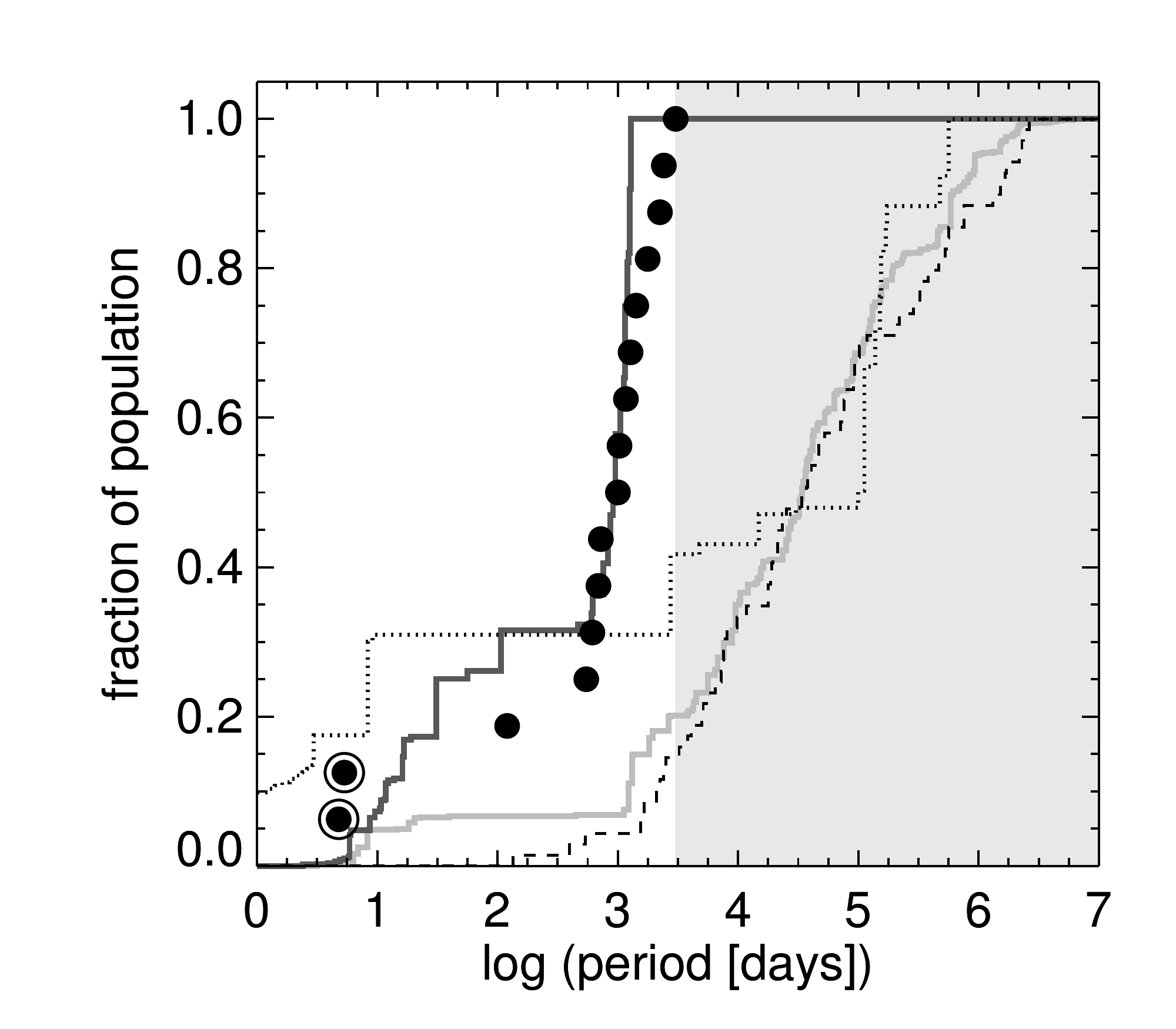} 
\caption{Cumulative distributions of orbital period for observed blue-straggler binaries in NGC 188 (black points) as compared to blue-straggler binaries in the NGC 188 model (mass transfer origin: solid dark gray line; collision origin: solid light gray line; merger origin: dotted line) and tertiaries from the field triple population (dashed line). We include model blue-straggler binaries of any period, and the gray-filled region marks periods beyond our observational completeness limit of 3000 days for orbital solutions. The two double-lined blue stragglers in NGC 188 are circled. Updated from \cite{GellerMathieu2012}.}\label{matfig9}
\end{figure}

\subsection{Implications for the Origins of the NGC 188 Blue Stragglers}
\label{matsec33}

The remarkable combination of comprehensive observations and sophisticated $N$-body modeling enable us to investigate the origins of the NGC 188 blue stragglers in great detail.  The binary properties of the blue stragglers, both frequency and distributions of orbital parameters and secondary masses, are keys to unlocking the mystery of blue straggler origins.  

Here we divide the discussion to focus first on the 14 NGC 188 blue stragglers in long-period ($>100$ day) binaries, all of which are single lined.  Then we discuss the two NGC 188 blue stragglers in short-period binaries, both of which are double lined.  Finally we briefly discuss the four NGC 188 blue stragglers that do not exhibit detectable radial-velocity variations.

\begin{figure}
\includegraphics[width=119mm]{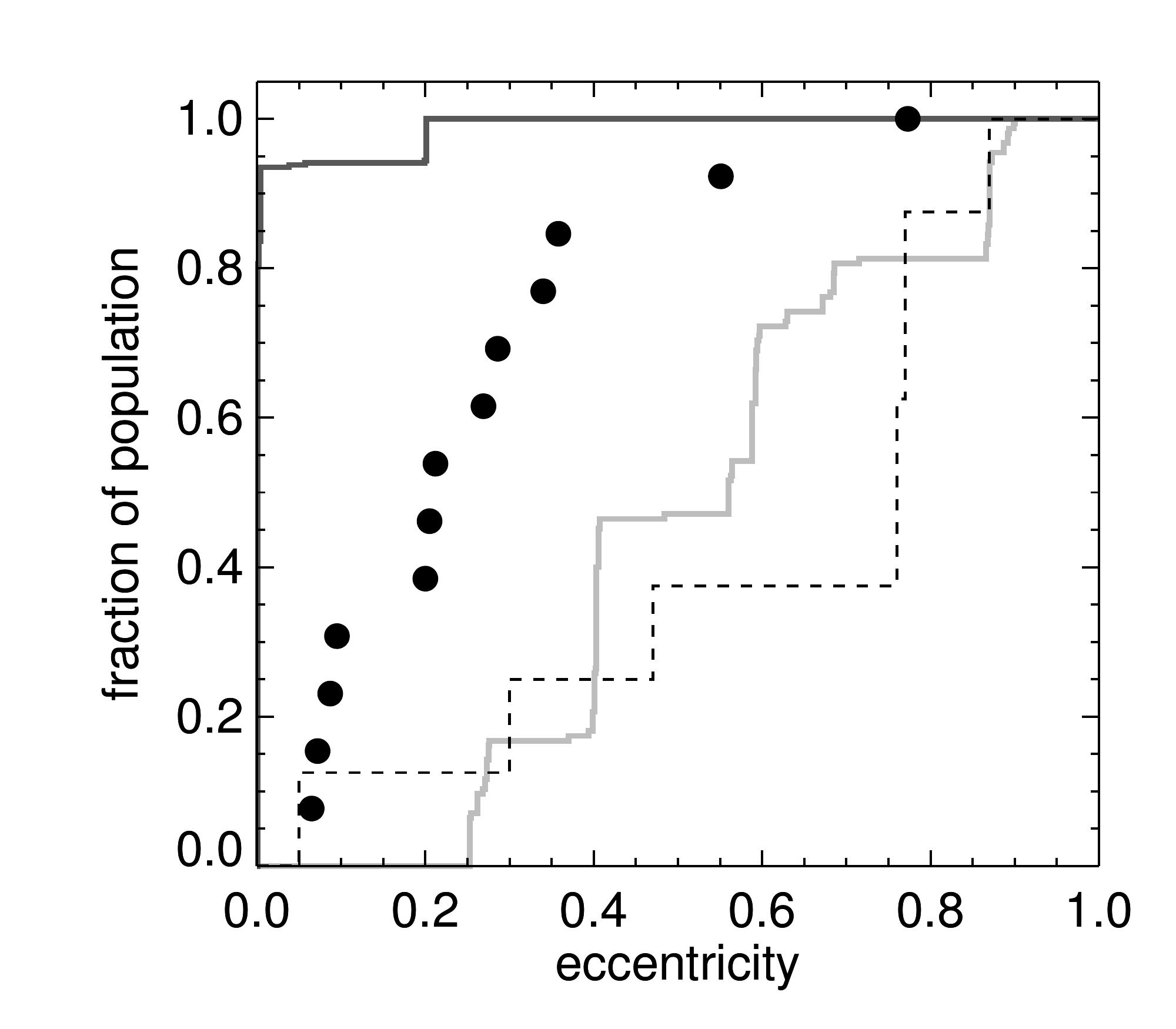} 
\caption{Cumulative distributions of orbital eccentricity\index{eccentricity} comparing the observed NGC 188 blue stragglers binaries with periods near 1000 days (black points) with predictions from the NGC 188 model for the collision and mass-transfer hypotheses, and the field triple population.  All line styles and symbols are the same as in Fig.~\ref{matfig9}.  Updated from \cite{GellerMathieu2012}.}\label{matfig10}
\end{figure}

\begin{figure}
\includegraphics[width=119mm]{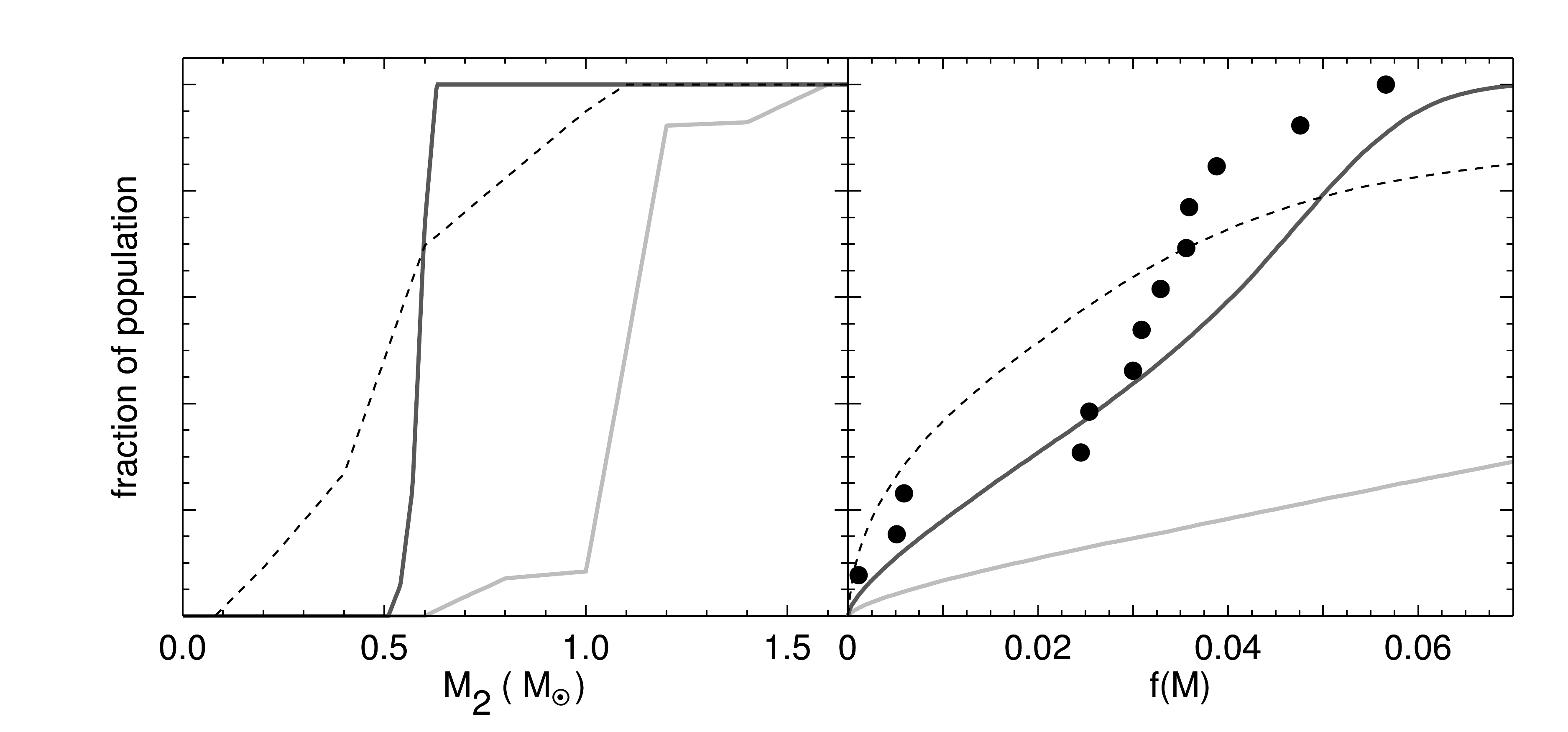} 
\caption{Left: Cumulative companion-mass distributions drawn from the NGC 188 model blue-straggler binaries with periods between 500 and 3000 days sorted by mass transfer origin (dark gray line) and collision origin (light gray line), and the evolved tertiary-mass distribution for field triples (dashed line; \cite{Tokovinin1997}, see Sect.~\ref{matsec343} for selection criteria). Right: Cumulative mass-function distributions, using the same line styles for the respective samples, and also including the observed mass functions for NGC 188 blue straggler in binaries with periods near 1000 days (black points). Updated from \cite{GellerMathieu2012}.}\label{matfig11}
\end{figure}

\subsubsection{The NGC 188 Blue Stragglers in Long-Period Binaries}
\label{matsec331}

Fourteen of the 20 blue stragglers in NGC 188 are found in long-period binaries. One of these blue stragglers has an orbital period near 100 days, and the rest have orbital periods near 1000 days. We first discuss the blue stragglers in 1000-day binaries.

In Fig.~\ref{matfig9}--\ref{matfig11}, we compare the observed binary properties of these NGC 188 blue stragglers to the predictions from the $N$-body model for blue stragglers in binaries formed by collisions, by mergers and by mass transfer. We did not include primordial triples in the NGC 188 model, and, although triples do form dynamically within the model, they are so few that the rate of KCTF blue-straggler formation is negligible.  Therefore the predictions for merger products from the NGC 188 model reflect blue stragglers produced by mergers within binaries. 

As can be seen in Fig.~\ref{matfig9}, the NGC 188 model predicts that mergers form essentially no blue stragglers in binaries with periods near 1000 days. Nearly all of the merger blue stragglers in the model that are within binaries with periods less than 3000 days are short-period contact systems that will eventually merge to form single blue stragglers. The vast majority of merger\index{merger} blue stragglers in wider binaries (mostly beyond our detection limit) formed through subsequent exchange encounters. The two populations are also distinguishable in orbital eccentricity; the very-short-period contact systems\index{contact binary} have zero eccentricities and the long-period exchange systems have high eccentricities.  

Only the collision and mass-transfer mechanisms produce blue stragglers in binaries with periods near 1000 days.  The predicted binary properties of blue stragglers formed by these two mechanisms are strikingly different.   

The collision mechanism predicts that 80\% of the companions to long-period blue stragglers are main-sequence stars, with most having have masses near that of the cluster turnoff (~1.1 $\Msun$).  This outcome reflects the tendency of dynamical encounters products to retain the most massive stars. Such companions are far more massive than observed for the NGC 188 long-period blue straggler binaries (Fig.~\ref{matfig3}). A Kolmogorov-Smirnov test rules out at $>99$\% confidence the hypothesis that the observed mass functions\index{binary mass function} are drawn from a parent population of collisional origin (Fig.~\ref{matfig11}).  Additionally, the collision mechanism predicts far too high a frequency of high-eccentricity orbits (Fig.~\ref{matfig10}). The eccentricity distribution for the blue stragglers formed by collisions in the $N$-body model is distinct from the observed distribution for the long-period NGC 188 blue stragglers with $>99$\% confidence \cite{GellerMathieu2011}. Therefore we rule out a primarily collisional origin for the NGC 188 blue stragglers in binaries with periods near 1000 days at very high confidence. 

All of the blue stragglers in the model that formed from mass transfer are members of binaries with periods less than 3000 days, securely detectable in our observations. Nearly all have white dwarf\index{white dwarf} companions with masses between 0.5 $\Msun$ and 0.6  $\Msun$ (Fig.~\ref{matfig11}). These predictions agree well with our observations of the NGC 188 blue stragglers in long-period binaries.  Most of the mass-transfer products in the NGC 188 model have periods near 1000 days and derive from Case C mass transfer. A smaller subset of the mass-transfer products from the model derive from Case B mass transfer and have helium white dwarf companions with shorter orbital periods.  These latter blue stragglers in the model may correspond to the one single-lined blue straggler in NGC 188 with an orbital period near 100 days and a mass function consistent with a low-mass companion.  

However, nearly all of the blue stragglers that formed by mass transfer in the NGC 188 model have circular orbits, while only three of the long-period NGC 188 blue stragglers have orbits consistent with being circular (the rest having a range of non-zero eccentricities). In {\tt NBODY6} (as well as many other modeling codes), tides are assumed to rapidly circularise the binary orbit during mass transfer. The NGC 188 model shows that dynamical encounters after the formation of blue stragglers by mass transfer are not sufficient to induce the number of observed eccentric blue straggler binaries. 
The three circular, long-period, single-lined NGC 188 blue stragglers are very likely candidates for the mass-transfer formation mechanism.  If mass transfer\index{mass transfer} is also responsible for the formation of the remaining long-period blue stragglers in NGC 188, then some form of ``eccentricity pumping««\index{eccentricity pumping} during the mass-transfer phase may be required. Indeed, recent theoretical work suggests that mass transfer will not always lead to circular orbits \cite{Soker2000,Bonacic2008,Sepinsky2009,LajoieSills2011}. Further development of these eccentricity pumping theories, and the inclusion of these mechanisms in models like {\tt NBODY6}, are highly desired.  

Furthermore, because of their isolation the field blue stragglers discussed by Preston in this volume (Chap. 4) are thought to derive primarily from mass transfer \cite{Carneyetal2005,Carneyetal2001}. Yet they also display non-zero eccentricities. Completion of our $HST$ survey for white dwarf companions to the NGC 188 blue stragglers will be very important for determining if the eccentric binaries in fact derive from mass transfer. For now, the eccentricity distribution predicted by the mass-transfer mechanism appears to be uncertain.?
A second discrepancy between the model predictions for the mass-transfer hypothesis and NGC 188 observations is that less than 10\% of the blue stragglers present in the model at the age of NGC 188 formed by mass transfer, while 70\% (14/20) of the NGC 188 blue stragglers are in long-period binaries. One hypothesis is that the efficiency of mass transfer is severely underestimated in the $N$-body model, and we discuss this in detail in Sect.~\ref{matsec342}.

We conclude that of the blue-straggler formation channels that are active in the NGC 188 model, only the mass-transfer mechanism produces blue straggler binaries whose orbital properties and secondary stars are closely consistent with the observed NGC 188 blue stragglers in long-period binaries.  However none of the formation channels, as implemented in {\tt NBODY6}, can fully explain both the observed frequency and the binary properties of these NGC 188 blue stragglers. 

\subsubsection{The NGC 188 Blue Stragglers in Short-Period Binaries}
\label{matsec332}

Both of the NGC 188 blue stragglers in short-period ($P < 10$ days) binaries are double-lined\index{double-lined spectroscopic binary} systems, one with a main-sequence companion near the cluster turnoff (WOCS ID 5078) and the other comprised of two blue stragglers (WOCS ID 7782).  The NGC 188 $N$-body model struggles to produce blue-straggler binaries with these characteristics.  Between 6 and 7.5 Gyr in the model, there are no blue stragglers in detached binaries with periods less than 10 days and companion masses greater than 0.9 $\Msun$. There are only two such systems present at earlier times in all 20 simulations.  Furthermore there are only three blue straggler -- blue straggler binaries in the NGC 188 model (out of 20 simulations) between 6 and 7.5 Gyr with short enough periods to have been detected as binaries.  These all have periods much greater than 10 days, and two were only bound for $<30$ Myr (one snapshot interval). Another four blue straggler -- blue straggler binaries are present in the model, but with periods beyond our observational detection limit.  All of these blue straggler -- blue straggler systems in the model were involved in exchange encounters that brought together into a binary two blue stragglers formed elsewhere.  Moreover, neither 5078 nor 7782 can be explained through isolated blue straggler formation mechanisms (like mass transfer, binary mergers or the KCTF mechanism). 

We conjecture that the evolutionary histories of these short-period NGC 188 blue-straggler binaries involve dynamical encounters\index{dynamical encounter}, despite being unable to produce similar systems in our NGC 188 model. Detailed observations of the binary properties of blue stragglers in additional open clusters are needed to determine whether such blue stragglers are typical or anomalous.  Of the few blue straggler populations already studied at this level of detail, there are two blue stragglers in short-period binaries in the field \cite{Carneyetal2001} and two blue stragglers in short-period binaries in M67\index{M67} \cite{Latham2007}.  The short-period field blue stragglers are both single lined with small mass functions, and so may not have massive companions (maintaining the possibility of Case B mass transfer, for example).  One of the short-period M67 blue stragglers is single-lined, and the other is S1082 (discussed in Sect.~\ref{matsec27} and \ref{matsec28}).  We note that the Hurley et al. \cite{Hurley2005} M67 model did not produce a system like S1082; indeed their model had no blue stragglers in higher-order systems at the age of M67 (although they only ran one simulation).  Thus both NGC 188 and M67 have blue stragglers in short-period binaries with massive companions, which $N$-body models\index{N-body simulation} do not readily produce.

\subsubsection{The Non-Radial-Velocity-Variable Blue Stragglers in NGC 188}
\label{matsec333}

There are four blue stragglers in NGC 188 that currently do not show significant radial-velocity variability.  In any given simulation within the NGC 188 model, we find on average five blue stragglers at the age of NGC 188 that do not have detectable binary companions (one of which, on average, is in a very wide binary, while the rest are single). The origins of these blue stragglers in the model are split roughly equally between mergers and collisions. 

The NGC 188 model predicts that blue stragglers produced by either mergers or collisions will be primarily single blue stragglers. About 9\% of the blue stragglers formed by mergers in the model have binary companions (4\% with orbital periods below our detection threshold of $10^4$ days).  About 12\% of the blue stragglers formed by collisions have orbital periods below our detection threshold, and only 33\% of the blue stragglers formed by collisions in the model have binary companions at any period.  Many of these single collision products formed with very wide companions that were quickly ionised by dynamical encounters.

Interestingly, nearly half of the blue stragglers that formed through collisions were members of dynamically formed hierarchical triple systems (often with a primordial binary as the inner system) during the snapshot interval prior to becoming a blue straggler.  The wider orbit of the tertiary companion increases the cross section for a stellar encounter, and the blue stragglers form through collisions during the consequent dynamical interactions. 

In short, the $N$-body model produces roughly the same number of non-radial-velocity-variable blue stragglers as is observed in the true cluster, and predicts that the majority of these are indeed single.

\begin{figure}
\includegraphics[width=119mm]{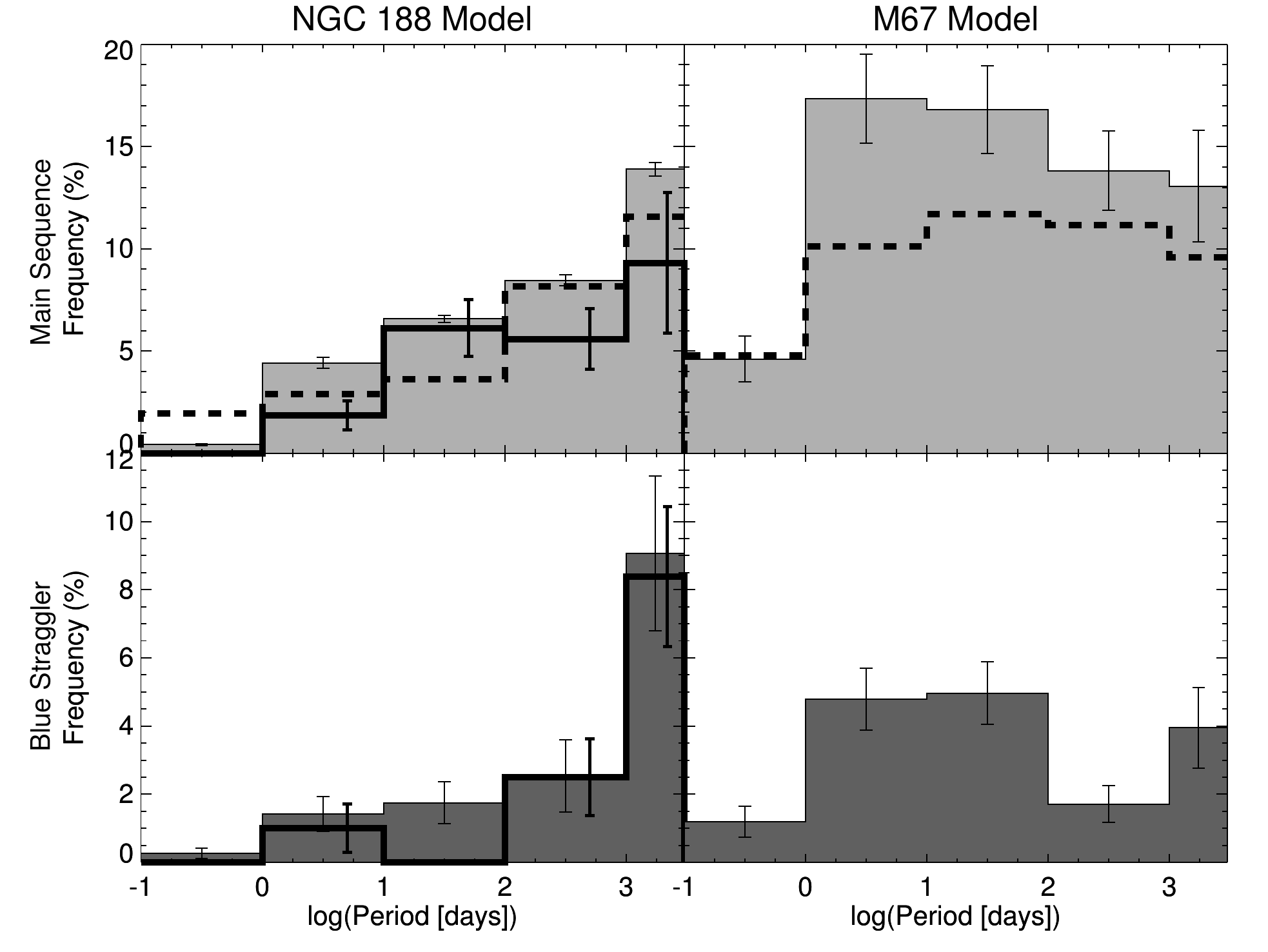} 
\caption{Comparison of the solar-type main sequence (top) and blue straggler (bottom) binary orbital period\index{orbital period} distributions in the Geller, Hurley \& Mathieu \cite{GellerHurleyMathieu2013} $N$-body model of NGC 188 at 7 Gyr (left) to those of the Hurley et al. \cite{Hurley2005} $N$-body model of M67 at 4 Gyr (right).  In the left panels, the solid black histograms show the observed NGC 188 main-sequence and blue-straggler binary orbital period distributions from \cite{GellerMathieu2012}; the main-sequence distribution is corrected for incompleteness. The dashed lines in the top panels show the initial distributions from the respective models. In all panels, the last bin is corrected for the smaller bin size.  The Hurley et al. model begins with a significantly larger frequency of short-period binaries than are observed in NGC 188 (or included in the NGC 188 model).  Adapted from \cite{GellerMathieu2012} and \cite{GellerHurleyMathieu2013}, and we refer the reader to these two references for similar comparisons of the observed and simulated eccentricity and companion-mass distributions.}\label{matfig12}
\end{figure}

\subsection{Outstanding Questions and Missing Pieces in the $N$-body Model }
\label{matsec34}

Despite closely reproducing observations of the solar-type main-sequence stellar population of the cluster, the NGC 188 model predicts significantly fewer blue stragglers at the age of NGC 188 than are observed. There are 20 blue stragglers in NGC 188, 16 of which are in binaries with periods $< 10^4$ days.  In contrast, the integrated blue straggler sample of the model counts only 6 blue stragglers at the age of NGC 188, one of which is in a detectable binary.  This deficit in blue straggler production rate indicates that the theory in the $N$-body model remains incomplete. We discuss here possible points for improvement.

The formation of blue stragglers rests on the cluster binary population. One hypothesis to explain the overall paucity of blue stragglers is that the choice of the initial binary population in the model is incorrect for NGC 188.  We discuss this in Sect.~\ref{matsec341}. 

When comparing the integrated blue-straggler population of the NGC 188 model to the observed NGC 188 blue stragglers, we find agreement in the shapes of the distributions of binary orbital periods (Fig.\ref{matfig12}), eccentricities and companion masses.  The agreement derives from the model distributions being dominated by blue stragglers formed by mass transfer; 60\% of the blue stragglers in binaries with periods $<3000$ days in the integrated sample of the model derive from the mass-transfer mechanism.  Therefore, a second hypothesis for the low number of blue stragglers is that the model underproduces blue stragglers through mass transfer specifically, which we discuss in Sect.~\ref{matsec342}.

Finally, the low frequency of dynamically formed triples in the NGC 188 model as compared to the observed frequencies of triples in the Galactic field and in a few open clusters suggest that NGC 188 may have formed with a population of primordial triples.  These triples, not included in the NGC 188 model, may contribute to blue straggler production, for example through the KCTF mechanism.  We discuss this hypothesis in Sect.~\ref{matsec343}.

\subsubsection{The Importance of an Accurate Initial Binary Population}
\label{matsec341}

The characteristics of the initial binary population\index{initial binary population} can have a strong impact on the production rates of blue stragglers. As a case in point, we compare the NGC 188 model to the Hurley et al. \cite{Hurley2005} model of M67 (created using {\tt NBODY4}, which is nearly identical to {\tt NBODY6}).  The Hurley et al. model produces 20 blue stragglers at the age of M67, roughly three times the number of blue stragglers produced by the NGC 188 model at the same age (despite the NGC 188 model having a larger total mass).  The difference between the blue straggler production rates in these two models is tied to their very different initial binary populations.    

The Hurley et al. model began with a 50\% binary frequency (integrated over all periods) and typical theoretical distributions of key binary properties: a uniform distribution in log periods and a thermal eccentricity distribution (which were subsequently evolved through the Kroupa \cite{Kroupa1995} pre-main-sequence evolution algorithm).  These initial distributions have a significantly higher frequency of short-period binaries than the NGC 188 model (Fig.~\ref{matfig12}).  

As a consequence, the frequency of main-sequence solar-type binaries with periods $<10^4$ days at 4 Gyr in the Hurley et al. model is 64\%, as compared to 28\% at 4 Gyr in the NGC 188 model and much higher than the binary frequency in NGC 188 itself (or M67; \cite{Latham2007}).  The overly high binary frequency in the Hurley et al. model is a direct result of its initially uniform distribution in $\log~P$ . We also find significant discrepancies between the distributions of binary orbital parameters in the Hurley et al. model and our observations of the NGC 188 binaries \cite{GellerMathieu2012}. The empirical initial conditions used in the NGC 188 model remedy these discrepancies.  
The higher frequency of short-period binaries increases blue straggler production through all mechanisms, enabling the Hurley et al. model to match the number of blue stragglers in M67\index{M67}. Interestingly, the relative rates of blue-straggler formation through the different mechanisms are roughly the same as in the NGC 188 model; 10\% of the blue stragglers in the M67 model at 4 Gyr formed through mass transfer, 40\% formed through collisions and 50\% formed through mergers. As a result, the frequency of detectable blue straggler binaries in the Hurley et al. model is only 30\%. Furthermore, of these detectable binaries a large fraction are found with short periods and high eccentricities, which is not observed in either NGC 188 or M67. 

Thus, although the Hurley et al. model produces roughly the correct number of blue stragglers found in NGC 188 and M67, the model fails to reproduce the blue-straggler (or main-sequence) binary properties as a result of the unrealistic initial binary population.  

On the other hand, a more realistic initial binary population in the NGC 188 model, which evolves to match the observed main-sequence binaries of the true cluster, does not reproduce the observed number of blue stragglers. We have attempted adjusting the parameters of the initial binary population and cluster central density, but we cannot simultaneously reproduce the number and binary properties of the NGC 188 blue stragglers, the binary proper-ties of the NGC 188 solar-type main-sequence stars, and the 7 Gyr cluster structure.

\subsection{Efficiency of Mass Transfer in the $N$-body Model}
\label{matsec342}

Our observations of the NGC 188 blue stragglers in long-period binaries --- which represent 70\% of the entire NGC 188 blue straggler population --- suggest that most, if not all, were formed by mass transfer\index{mass transfer}.  Yet, the mass-transfer mechanism is woefully inefficient at producing blue stragglers in the $N$-body model.  On average, less than one blue straggler at the age of NGC 188 is produced by mass transfer in the model. Even evolving the initial binary population in isolation (disregarding stellar encounters or escape from the cluster, both of which tend to decrease the number of blue stragglers in the cluster formed by mass transfer), we cannot produce the 14 NGC 188 blue stragglers in long-period binaries through the mass-transfer mechanism.  If mass transfer is indeed responsible for producing these NGC 188 blue stragglers in long-period binaries, then the physics of mass transfer in {\tt NBODY6} (and models using similar binary evolution prescriptions) is not accurate. 

As described in Section~\ref{matsec321} (and \cite{Hurley2002}), during RLOF {\tt NBODY6} compares a critical mass ratio, $q_c$, to the binary mass ratio, $q_1$, to determine whether a given binary will go through mass transfer on a thermal or nuclear timescale (i.e., stable), or a dynamical timescale (i.e., unstable).  Moreover, for a given binary undergoing RLOF, the parameter $q_c$ determines whether a blue straggler can be created or whether the system will go through a common-envelope episode and not produce a blue straggler.  Thus in terms of creating blue stragglers through mass transfer, $q_c$ is a key parameter in the model. 

However, the value of $q_c$ for a given binary is highly uncertain and depends on how conservative is the mass transfer.  For a given binary, different developments in the literature (e.g, \cite{HjellmingWebbink1987,Hurley2002,ChenHan2008a,ChenHan2008b, Belczynsky2008}) provide different $q_c$ values (by a factor of a few).  There is also growing theoretical evidence that this standard $q_c$ parameterization may not match predictions from more detailed binary mass-transfer models (\cite{WoodIvanova2011,Passyetal2012}; see also Chap. 8).  

In the NGC 188 model, we employ the Hjellming \& Webbink \cite{HjellmingWebbink1987} formula and do not account for non-conservative mass transfer explicitly in the calculation of $q_c$. The result of this method is to send the vast majority of possible proto-blue-straggler systems undergoing RLOF through the common-envelope channel, and thereby to produce a population of main-sequence Ð white dwarf binaries in circular orbits with periods of a few hundreds to thousands of days. (See Fig. 9 in \cite{GellerHurleyMathieu2013}.) Such long-period circular binaries are not observed amongst the true solar-type main-sequence binary population in NGC 188 (nor in the Galactic field or any other open cluster binary population).  

We suggest that these systems may, in reality, undergo stable mass transfer to produce blue stragglers.  On average, there are 9 such binaries in the model at the age of NGC 188 that perhaps should instead have undergone Case C (or B) mass transfer\index{Cases A, B, C of mass transfer}, and thereby become blue stragglers with white dwarf companions orbiting at long periods. If these spurious common-envelope products were converted to mass-transfer blue stragglers, the result-ing frequency and binary properties of the NGC 188 model blue stragglers would nearly reproduce the blue straggler population of the true cluster. These specific systems in the NGC 188 model are prime candidates for follow-up through detailed binary evolution models to determine whether indeed they should go through stable or unstable mass transfer.    
 
In summary, the combination of (1) the deficit of blue stragglers formed by mass transfer, compared to the 14 NGC 188 blue stragglers with binary properties indicative of a mass transfer origin, and (2) the excess of spurious long-period circular solar-type main-sequence Ð white-dwarf binaries suggests that the physics of mass transfer as incorporated in the $N$-body model is not yet accurate. The efficiency of stable mass transfer may be significantly underestimated in {\tt NBODY6} and in other codes that use similar binary evolution procedures.

\subsubsection{Primordial Triples and the KCTF Formation Channel}
\label{matsec343}

An alternative hypothesis to explain the paucity of blue stragglers in the NGC 188 model is that the model is missing a population of primordial hierarchical triples.  We did not include primordial triples in the NGC 188 model, although triples do form dynamically and their subsequent dynamical evolution is modeled in detail.  However, the frequency of dynamically formed triples is never high enough to reproduce the observed frequencies of triples in the Galactic field or in the few open clusters where such data are available.  Although we know very little about the triple population\index{triple system} of NGC 188, if the cluster currently has a similar frequency of triples as observed for solar-type stars in the field, then our $N$-body model indicates that the cluster must have formed with these triples in its primordial population.  

Because triples may be important for blue straggler production through the KCTF mechanism and through collisions resulting from dynamical encounters, we performed a few additional models of the cluster with increased numbers of primordial triples  \cite{GellerHurleyMathieu2013}.  We included ad hoc triple populations, associating solar-type main-sequence binaries having periods between 2 and 50 days with solar-type tertiaries at periods of 700--3000 days in an attempt to maximise the contribution of the KCTF mechanism in particular to blue stragglers at periods near 1000 days at seven gigayears.  

Even with this ``optimised'' triple population, we find that including primordial triples does not dramatically increase the number of blue stragglers at the age of NGC 188.  Placing every solar-type main-sequence binary with an orbital period between 2 and 50 days in a triple (roughly 200 systems in total) on average only results in a total of 7--8 blue stragglers at the age of NGC 188 (as compared to an average of 6 blue stragglers in the NGC 188 model with-out primordial triples).  

However, the inclusion of primordial triples does dramatically increase the detectable blue-straggler binary frequency, up to 80\% for our model with 200 triples. The majority of these additional blue-straggler binaries formed the blue stragglers through collisions during stellar encounters involving primordial triples, after which the newly formed blue stragglers managed to retain companions from the encounter at detectable orbital periods. Thus triples may in fact play an important role in producing blue-straggler binaries through collisions (as was suggested by \cite{MathieuGeller2009,LeighSills2011}). 

Only about 25\% of these detectable blue-straggler binaries formed in relative isolation through Kozai-induced processes.  It is possible that the Mardling \& Aarseth \cite{Mardling2001} analytic development for modeling Kozai oscillations\index{Kozai cycle} and tidal processes in triple stars, which is employed in {\tt NBODY6}, may underestimate the efficiency of KCTF blue straggler production.  Further investigation into this possibility is needed, as are additional $N$-body models with more varied populations of primordial triples --- ideally based directly on observations of triple populations in open clusters. At least in our current NGC 188 $N$-body simulations with {\tt NBODY6}, the KCTF mechanism is not a significant contributor to blue straggler production.  

In the absence of a secure treatment in {\tt NBODY6}, here we explore the KCTF theoretical prediction that blue-straggler companions should derive from the primordial tertiary population. Specifically, we compare the orbital parameters and masses for the long-period NGC 188 blue straggler binaries to those of observed triple stars. 
Very little is currently known about the triple populations in open clusters. As a proxy, we compare with the triples observed in the solar neighbourhood from the Multiple Star Catalogue (MSC; \cite{Tokovinin1997}).  The MSC is not a complete sample, but is representative of field triples and is currently the largest sample of triples available in the literature.  We select from the MSC all triples that have inner binaries with masses and orbital parameters that could conceivably merge to form blue stragglers similar to those in NGC 188.  Specifically we select triples with inner binaries of total masses between 1.2 $\Msun$ and 2.2 $\Msun$ (i.e., from just above the turnoff mass to twice the turnoff mass of NGC 188), inner orbital periods less than 10 days, and outer orbital periods less than 3000 days.  

The outer-period distribution for this sample of triples is consistent with the periods of the long-period NGC 188 blue stragglers (although the majority of triples have much larger periods; see Fig.~\ref{matfig9}). A subset of eight triples from this sample also have eccentricity measurements, shown in Fig.~\ref{matfig10}. Their eccentricity distribution is shifted to larger eccentricities than observed for the NGC 188 blue stragglers, at the 97\% confidence level. 

The distributions of companion masses for this sample of MSC triples are shown in Fig. ~\ref{matfig3} and ~\ref{matfig11}, evolved to the age of NGC 188.  Geller \& Mathieu \cite{GellerMathieu2011} show that if we draw companions to the NGC 188 blue stragglers from the MSC tertiary mass distribution, there is only a 1.8\% probability that they would both reproduce the observed mass function distribution of the NGC 188 blue stragglers in long-period binaries and result in all single-lined systems. Thus both the companion masses and eccentricities of the NGC 188 blue stragglers are inconsistent with those of the tertiaries of solar-type field triple systems in the MSC.

In summary, the KCTF channel encounters difficulties in explaining the observed properties of the NGC 188 blue stragglers in long-period binaries.  The KCTF mechanism does not operate efficiently in our $N$-body models that include large populations of primordial triples, producing far less than the 14 observed NGC 188 blue stragglers in long-period binaries.  Instead triples in our models contribute to blue straggler formation primarily through collisions resulting from stellar encounters.  It is possible that this low KCTF efficiency is a failing of the implementation of KCTF in {\tt NBODY6}.  However, potential KCTF progenitors in the observed triple population of the solar neighborhood also predict companions that are too massive and orbits that are too eccentric to explain the binary orbital properties of the NGC 188 blue stragglers in long-period binaries. 

\subsubsection{Bimodal Blue-Straggler Spatial Distributions}
\label{matsec344}

As discussed in Section~\ref{matsec26}, NGC 188 has a bimodal blue-straggler radial distribution\index{radial distribution} reminiscent of those observed in many globular clusters.  The NGC 188 model does not show this same structure in the blue-straggler radial distribution.  At the age of NGC 188, the blue stragglers in the model are centrally concentrated, and there is not a distinct halo population like that observed in the true cluster.  Only 17\% of the simulated blue stragglers are found outside of 5 pc at the age of NGC 188, as compared to 30\% in the true cluster. 

This lack of a detectable bimodal radial distribution for the blue stragglers in the NGC 188 model may correspond with the overall paucity of blue stragglers discussed above, in that the model appears to underproduce blue stragglers through non-dynamical formation channels (e.g., mass transfer or the KCTF mechanism). Blue stragglers formed through such channels may form throughout the cluster, including the halo. This result should be revisited when the model more accurately reproduces the cluster blue straggler number.

Interestingly, though, the binary frequencies and distributions of orbital parameters for the NGC 188 core and halo blue stragglers are statistically indistinguishable.  This finding suggests that the bimodal blue straggler radial distribution seen in NGC 188 is not the result of distinct formation channels.

\subsection{Summary of Findings from $N$-body Modeling of NGC 188}
\label{matsec35}

Our NGC 188 $N$-body model\index{N-body simulation} provides detailed predictions for the binary properties of blue stragglers formed by different formation processes. As is often the case, when one looks in greater detail at a problem, both answers and questions arise. We will focus first on the predictions from the model that appear most secure.  Specifically, the NGC 188 $N$-body model predicts that within a rich open cluster after several gigayears of dynamical evolution:
\begin{itemize}
\item Blue stragglers produced by mass transfer will have a very high binary frequency (100\% in the NGC 188 model, all with orbital periods $<10^4$ days and most near 1000 days).  Except for the very few mass-transfer blue stragglers that undergo exchange encounters, these blue stragglers are predicted to have white-dwarf companions (the remnant cores of the donor stars).  Most of these mass-transfer blue stragglers form by Case C mass transfer and have CO white-dwarf companions that have masses be-tween roughly 0.5 $\Msun$ and 0.6 $\Msun$; a smaller subset form through Case B mass transfer and have He WD companions with masses below 0.5 $\Msun$. 
\item Blue stragglers produced by collisions will have a low binary frequency (12\% with orbital periods$<10^4$ days, and 33\% over all periods), and those that are in binaries with orbital periods around 1000 days tend to have high eccentricities and companions with masses near that of the cluster turnoff.
\item Blue stragglers produced by binary mergers will also have a low binary frequency (4\% with orbital periods $<10^4$ days, and 9\% over all periods), and those that are in binaries are either in short-period contact systems on their way towards merging or in long-period eccentric binaries produced by exchange encounters.  Essentially no blue stragglers produced by mergers in binaries will be found with periods near 1000 days.
\end{itemize}

Our comparisons between these predictions and our observations of the NGC 188 blue stragglers rule out at very high confidence an origin through collisions\index{collision} for the NGC 188 blue stragglers in binaries with periods near 1000 days (which comprise 70\% of the NGC 188 blue stragglers).  Mergers strictly within binaries also cannot produce blue stragglers in long-period binaries.  Only the mass-transfer and KCTF mechanisms naturally produce blue stragglers in binaries with periods near 1000 days. 

The mass-transfer predictions for the secondary-star masses are closely consistent with the observed companion mass distributions. The recently detected white dwarf companions for some of the NGC 188 blue stragglers also agree with model predictions for an origin through mass transfer.

However, the mass-transfer production rate in the model was too low to reproduce accurately the number of blue stragglers.  We suggest this reflects on the implemented mass-transfer theory, and in particular that detailed binary models are necessary to define the boundary between stable and unstable mass transfer. The orbital eccentricity distribution predicted by the mass-transfer mechanism also remains a key open question.
The NGC 188 blue stragglers in short-period binaries were most likely involved in previous close stellar encounters. We suspect that some or all of these blue stragglers were formed prior to their current binary configurations. However, the NGC 188 model struggles to produce blue stragglers with these binary characteristics.  
  
Finally, the model does predict the same number of single (or very-long-period binary) blue stragglers as we observe in NGC 188.  Roughly half of these blue stragglers in the model came from collisions with the remainder from binary mergers. 
  
We conclude that the binary properties of the NGC 188 blue stragglers in long-period binaries are most closely reproduced by the mass-transfer mechanism.  However, a combination of formation channels (indeed, perhaps all channels) are likely required to reproduce the full blue straggler population of NGC 188.

\section{Conclusions}
\label{matsec4}

From a dynamical perspective, open clusters represent valuable laboratories that are intermediate between the high-density cores of many globular clusters and the collisionless environment of the Galactic field.

The recent deep mining of the open cluster NGC 188, and before that M67, yield a rich description of blue stragglers in old clusters, as summarised in Sect.~\ref{matsec29}. The overarching conclusion from this observational work is that open cluster blue stragglers are primarily binary systems, with the strong implication that their origins are inextricably linked to the rich primordial bi-nary populations of open clusters.

The predominance of orbital periods near 1000 days combined with a secondary mass distribution narrowly peaked at white-dwarf-like masses, as well as recent detections of white dwarf companions among several NGC 188 blue stragglers, point to most blue stragglers in old open clusters forming through mass transfer in binaries having asymptotic branch primary stars. Several remarkable short-period double-lined binaries point to subsequent dynamical exchange encounters modifying the blue straggler population after formation, and provide at least one example of a likely collisional origin for a blue straggler.

The observations open critical questions and new opportunities. The origin of the spin angular momentum of blue stragglers is unknown; the modest rotation rates are neither near break-up velocities as some theories predict nor like those of similar effective-temperature stars at ages of several gigayears. The few dynamical mass measurements find the blue stragglers (in short-period binaries) to be underluminous compared to normal evolutionary tracks, in one instance by as much as a factor of a few. Finally, the white-dwarf detections indicate effective temperatures that correspond to ages of several hundred megayears. In contrast to the age of NGC 188, these blue stragglers were formed ``yesterday''. With determination of the white-dwarf masses and thus the evolutionary states of the donor stars, these blue stragglers provide opportunities for detailed modeling of their mass transfer origins.

Extensive $N$-body modeling of NGC 188 with empirical initial conditions is able to reproduce the properties of the cluster, including the main-sequence solar-type binary population that is so critical to blue-straggler formation rates. The current models also reproduce well the binary orbital properties of the blue stragglers. However, they fall well short of producing the observed number of blue stragglers or their very high binary frequency. 

Again, this challenge is at the same time an opportunity to improve essential physics within stellar and cluster evolution. In the case of the low blue-straggler formation rate in the current models, our analyses suggest that this is the result of an inaccurate parameterization of red-giant and asymptotic-giant mass transfer. This feature of the model also produces an excess of main-sequence -- white-dwarf long-period binaries not observed in the cluster, with numbers comparable to the deficit of blue stragglers. Demanding that the main-sequence stars in these binaries actually become blue stragglers informs the relative rates of common-envelope versus mass-transfer evolution.

The progenitors of these spurious main-sequence -- white-dwarf long-period binaries from the NGC 188 model offer ideal test cases for detailed bi-nary evolution models investigating the stability of mass transfer and blue straggler production.  Furthermore, with increasing computational speeds and recent software advances (like MESA\index{MESA code}; \cite{Paxtonetal2011}, and AMUSE\index{AMUSE code}; \cite{McMillanetal2012}), future $N$-body star cluster simulations may include ``live«« binary models for systems undergoing RLOF, and perhaps ``live«« collision models as well, thereby avoiding many of the assumptions necessary for the parameterised models we use today.  

Still, within the current $N$-body framework, we cannot yet simultaneously reproduce all of the observed properties of the NGC 188 blue stragglers along with the cluster mass, central density, radial structure, main-sequence binary population, etc.  Thus the challenge remains: to reproduce the observed NGC 188 blue straggler population within a coherent $N$-body model of the cluster.  

\pagebreak

\backmatter
\printindex


\end{document}